\documentclass{CUP-JNL-DTM}%

\usepackage{graphicx}
\usepackage{multicol,multirow}
\usepackage{amsmath,amssymb,amsfonts}
\usepackage{mathrsfs}
\usepackage{amsthm}
\usepackage{rotating}
\usepackage{appendix}
\usepackage{natbib}
\usepackage{ifpdf}
\usepackage[T1]{fontenc}
\usepackage{newtxtext}
\usepackage{newtxmath}
\usepackage{textcomp}
\usepackage{xcolor}
\usepackage{lipsum}
\usepackage[colorlinks,allcolors=blue]{hyperref}
\usepackage{soul, xcolor}

\theoremstyle{definition}

\numberwithin{equation}{section}

\usepackage{subcaption}
\usepackage{float}
\usepackage{color} 
\usepackage{tablefootnote}
\usepackage[bottom]{footmisc}
\usepackage[super]{nth}
\usepackage{framed}
\usepackage{array}
\usepackage{booktabs}
\usepackage{verbatim}
\usepackage{eurosym}
\usepackage[all]{nowidow}
\usepackage{commath}
\usepackage{listings}
\usepackage[normalem]{ulem}
\useunder{\uline}{\ul}{}
\usepackage{todonotes}
\usepackage{bm}
\usepackage{soul}

\newcolumntype{L}{>{\arraybackslash}m{5cm}}

\jname{Data/Math}
\articletype{RESEARCH ARTICLE}
\jyear{2023}

\begin{document}

\begin{Frontmatter}

\title[Article Title]{Bayesian system identification for structures considering spatial and temporal correlation}

\author[1,2]{Ioannis Koune}{\orcid{0000-0003-3459-9165}}
\author[2]{Arpad Rozsas}\orcid{0000-0003-2679-1384}
\author[2]{Arthur Slobbe}\orcid{0000-0002-1634-7960}
\author[1]{Alice Cicirello}\orcid{0000-0002-6556-2149}

\authormark{Koune \textit{et al}.}

\address[1]{\orgdiv{Faculty of Civil Engineering and Geosciences}, \orgname{Delft University of Technology}, \orgaddress{\city{Delft}, \postcode{2628 CN}, \state{South Holland},  \country{Netherlands}}}

\address[2]{\orgdiv{Building, Infrastructure and Maritime Unit}, \orgname{TNO}, \orgaddress{\city{Delft}, \postcode{2629 JD}, \state{South Holland},  \country{Netherlands}}}

\address{\textbf{Corresponding author:} Ioannis Koune; \email{i.c.koune@tudelft.nl}}

\keywords{Structural health monitoring, Bayesian system identification, parameter estimation, model prediction uncertainty}


\abstract{

The decreasing cost and improved sensor and monitoring system technology (e.g. fiber optics and strain gauges) have led to more measurements in close proximity to each other. When using such spatially dense measurement data in Bayesian system identification strategies, the correlation in the model prediction error can become significant. The widely adopted assumption of uncorrelated Gaussian error may lead to inaccurate parameter estimation and overconfident predictions, which may lead to sub-optimal decisions. This paper addresses the challenges of performing Bayesian system identification for structures when large datasets are used, considering both spatial and temporal dependencies in the model uncertainty. We present an approach to efficiently evaluate the log-likelihood function, and we utilize nested sampling to compute the evidence for Bayesian model selection. The approach is first demonstrated on a synthetic case and then applied to a (measured) real-world steel bridge. The results show that the assumption of dependence in the model prediction uncertainties is decisively supported by the data. The proposed developments enable the use of large datasets and accounting for the dependency when performing Bayesian system identification, even when a relatively large number of uncertain parameters is inferred.

}

\end{Frontmatter}

\section*{Impact Statement}
In Bayesian system identification for structures, simplistic probabilistic models are typically used to describe the discrepancies between measurement and model predictions, which are often defined as independent and identically distributed Gaussian random variables. This assumption can be unrealistic for real-world problems, potentially resulting in underestimation of the uncertainties and overconfident predictions. We demonstrate that in a real-world case study of a twin-girder steel bridge, the inclusion of correlation is decisively favoured by the data. In the proposed approach, both the functional form of the probabilistic model and the posterior distribution over the uncertain parameters of the probabilistic model are inferred from the data. A novel efficient log-likelihood evaluation method is proposed to reduce the computational cost of the inference.



\setstcolor{red}

\section{Introduction}
\label{chapter:intro}


\subsection{Motivation}



Structural Health Monitoring (SHM) methods based on probabilistic approaches have seen significant development in recent years \citep{Farrar2012} and have been applied for system identification and damage detection for various types of structures including bridges \citep{Behmanesh2014}, rail \citep{Lam2014}, offshore oil and gas installations \citep{Brownjohn2007}, offshore wind farms \citep{Rogers2018} and other civil engineering structures \citep{Chen2018}. The Bayesian system identification framework established in \cite{Beck1998} uses measurements of structural responses obtained from sensors in combination with computational physics models to infer uncertain parameters, calibrate models, identify structural damage and provide insight into the structural behaviour \citep{Huang2019}. In Bayesian statistics the problem is cast as a parameter estimation and model selection problem, often referred to as system identification in the SHM literature \citep{Katafygiotis1998}. Specifically, previous knowledge about the system parameters to be inferred is represented by statistical distributions and combined with measurements to infer the posterior parameter distribution. A key advantage of this approach is that it provides a rigorous framework for combining prior knowledge and data with a probabilistic description of the uncertainties to obtain a posterior distribution over non-directly observed parameters of interest (the so-called latent variables) using directly observed responses. For example, the rotational stiffness of a support can be estimated based on measured deflections \citep{Ching2006,Lam2018}.

In parallel with the probabilistic methods for SHM, sensor and monitoring technologies have seen significant progress in recent years. These technologies can provide higher accuracy and improved measurement capabilities, e.g. by utilizing fiber optic strain sensors \citep{Barrias2016,Ye2014}. Fiber optic strain sensors provide measurements with high spatial and temporal resolution as large numbers of sensors with high sampling rates are used in the same structure. System identification is carried out under the assumption that there is sufficient information in the measurements, so that the data can overrule the prior assumption on the latent variables. Therefore, utilizing the additional information contained in these measurements can potentially improve the accuracy of our predictions, reduce the uncertainty on the inferred system parameters, and lead to improved physical models that can more accurately capture the structural behaviour. However, when using measurements from dense sensor layouts, such as fiber optic strain sensors, the discrepancies between model prediction and observations are expected to be dependent. This dependence has to be considered in the system identification to avoid inaccurate parameter estimation and overconfidence in the model predictions.

\subsection{Problem statement}
\label{section:problem_statement}

\begin{figure}[H]
    \centering
    \FIG{\includegraphics[width=0.70\textwidth]{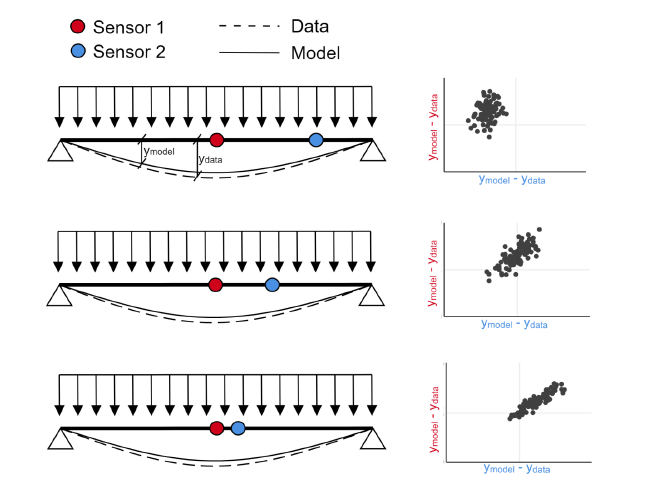}}
    \caption{Illustration of the impact of correlation in the model prediction error for the fictitious case of a simply supported beam with two sensors.}
    \label{fig:impact_of_correlation}
\end{figure}

Current approaches in Bayesian inference for structures largely neglect the dependencies in the model prediction error. Instead, it is typically assumed that the prediction error is Gaussian white noise, i.e. uncorrelated with zero mean \citep{Lye2021}. When using closely spaced measurements and model predictions, e.g. in the case of time series with high sampling rates or spatial data from densely spaced sensors, dependencies may be present in the model prediction errors \citep{Simoen1998}. The strength of the correlation typically depends on the proximity of the measurements in time and the spacing of sensors on the structure. A fictitious example of a simply supported beam where the error between measurement  and model prediction for two sensors is modeled as a bivariate Normal distribution, explicitly accounting for the spatial correlation of three different sets of measurements, is shown in \autoref{fig:impact_of_correlation} for illustration purposes. Disregarding the spatial and temporal measurements correlation, by  enforcing the assumption of independence can lead to large errors in the posterior distribution of the inferred parameters, as correlation has been shown to have an impact on the information content of measurements \citep{Papadimitriou2012}, the maximum likelihood and maximum a posteriori estimates of the parameters of interest, and the posterior uncertainty \citep{Simoen2013}.  

To consider correlations in Bayesian system identification for structures poses a number of challenges for the modeler. An appropriate functional form of the prediction error correlation is not known a priori, and due to the prevalence of the independence assumption there is limited information available on how to model it. Additionally, it is not known to what degree the correlation is problem specific. Hence, to perform Bayesian system identification on real-world structures when spatial and/or temporal dependence are present, we identified the following open issues: 

\begin{itemize}
    \item[1.] Appropriate models for the spatial and temporal correlations must be included in the probabilistic model that describes the uncertainties.
    \item[2.] Bayesian inference must be performed in a computationally efficient manner when large datasets and combined spatial and temporal dependencies are considered.
\end{itemize}

\subsection{Approach}
\label{section:approach}

\begin{figure}[htb!]
    \centering
    \FIG{\includegraphics[width=0.8\textwidth]{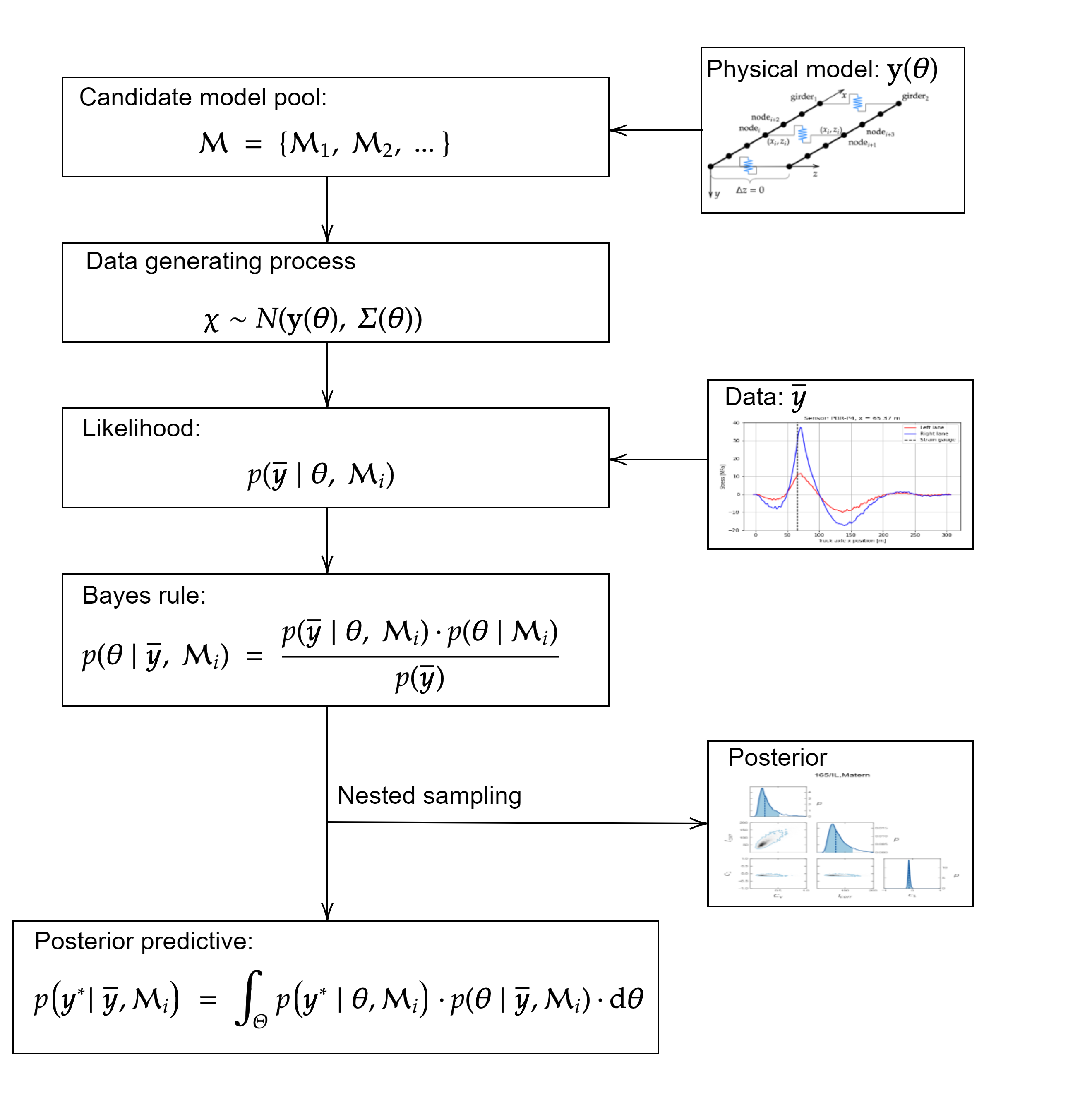}}
    \caption{Overview of the Bayesian inference approach used in this work.}
    \label{fig:overview_approach}
\end{figure}

The approach proposed in this paper to address the issues mentioned above can be summarized as follows. First, a mathematical model of the data generating process is formulated. This model is composed of a physical model describing the response of the structure, and a probabilistic model describing the measurement and model prediction error including the spatial and temporal correlation. Both the measurement and model prediction error are taken as normally distributed, and the strength of the correlation is assumed to be dependent on the distance between measurements (in time and/or in space). This dependence is modelled by a set of kernel functions. A pool of candidate models is defined, with each model considering a different kernel function to describe the correlation in the physical model prediction error. Bayesian inference is performed to obtain the posterior distribution of physical and probabilistic model parameters based on the data. The posterior probability and Bayes factor are calculated for each candidate model, making it possible to evaluate how strongly a given model is supported relative to the other candidate models based on the data. The proposed approach is illustrated in \autoref{fig:overview_approach}. More details on the individual building blocks are given in Section \ref{section:methods_and_tools}.

Second, a strategy is presented for performing system identification for relatively large datasets ($N > 10^2$ for temporal dependencies and $N > 10^3$ for combined spatial and temporal dependencies) by efficiently evaluating the log-likelihood and the evidence. We propose a procedure for exact and efficient log-likelihood calculation by (i) assuming separability of the spatial and temporal correlation \citep{Genton2007}; (ii) exploiting the Markov property of the Exponential kernel \citep{Marcotte2018}; and (iii) using the nested sampling strategy \citep{Skilling2006} to reduce the computational cost of estimating the evidence under each model. The accuracy of the proposed approach is initially investigated on a case study using synthetic data, and subsequently the feasibility of the approach for its use in real-world cases is demonstrated through a twin-girder steel road bridge case study. In the real-world use case, stress influence lines obtained from controlled load tests are used to estimate the posterior distribution of a set of uncertain, unobservable parameters. The accuracy and uncertainty of the posterior predictive stress distributions obtained from each candidate model are compared, to determine the benefit of using a larger dataset and considering dependencies.

\section{Previous work}
\label{section:literature}



In the Bayesian system identification literature, it is typically assumed that the prediction error is Gaussian white noise, i.e. uncorrelated with zero mean (\cite{Pasquier2015}, \cite{Chiachio2015}, \cite{Astroza2017}, \cite{Mthembu2011}). In some studies, e.g. \cite{Ebrahimian2018} and \cite{Goller2011} the variance of the model prediction error is included in the vector of inferred parameters, however, dependencies are not considered. In other works, such as \cite{Simoen2015}, \cite{Pasquier2020} and \cite{Vereecken2022} the parameters that define the uncertainty (with or without considering dependencies) are estimated using a subset of the available data. This approach, however, results in the use of data for inferring nuisance parameters and may not be practical when limited data is available. Examples of inference of the uncertainty parameters can be found in applications outside of structural engineering, e.g. in geostatistics \citep{Diggle2002}. 

To the best of the authors knowledge, \cite{Simoen2013} is the only work concerning model prediction error correlation in Bayesian system identification for structures, and investigates the impact of considering dependencies in model prediction error in Bayesian system identification. The aforementioned study presents an approach with many similarities to the proposed one. Bayesian inference and model selection is applied to a pool of candidate models to infer the distribution of uncertain parameters and to determine the strength of the evidence in favour of each model. The physical and probabilistic parameters are inferred for a simple linear regression example as well as a reinforced concrete beam example using modal data. In both cases the datasets are composed of synthetic observations polluted with correlated noise. Furthermore, the posterior distributions are assumed to be Gaussian, allowing for a computationally efficient asymptotic approximation to be utilized to obtain the posterior and evidence.

We focus on the feasibility of the approach in a practical application with real-world data, where the ground truth of the correlation structure and parameters are not known and the posterior and evidence are not approximated analytically. We instead utilize nested sampling to estimate the evidence, ensuring the applicability of the approach in cases where the Gaussian assumption for the posterior is not valid. Additionally, we address the issue of efficiently calculating the log-likelihood for large datasets with combined spatial and temporal dependencies under the assumption of separable space and time covariance.

\section{Methods and tools}
\label{section:methods_and_tools}

\subsection{Continuous Bayes theorem}
\label{section:continuous_bayes}
The Bayes theorem of conditional probability for continuous random variables can be written as \citep{Gelman2013}:

\begin{equation}
    \label{eq:bayes_cont}
    p(\bm{\theta}|\overline{\bm{y}},\mathcal{M}) = \frac{p(\overline{\bm{y}}|\bm{\theta}, \mathcal{M}) \cdot p(\bm{\theta}|\mathcal{M})}{\int_{\Theta}p(\overline{\bm{y}}|\bm{\theta}, \mathcal{M}) \cdot p(\bm{\theta}|\mathcal{M}) \cdot \mathrm{d}\bm{\theta}}
\end{equation}

\noindent
where:

\begin{itemize}
    \item $\bm{\theta}$ is a vector of uncertain parameters;
    \item $\overline{\bm{y}}$ a vector of observations;
    \item $\mathcal{M}$ denotes the model;
    \item $p(\bm{\theta}| \overline{\bm{y}}, \mathcal{M})$ is the posterior distribution;
    \item $p(\overline{\bm{y}}|\bm{\theta}, \mathcal{M})$ is the likelihood;
    \item $p(\bm{\theta}|\mathcal{M})$ is the prior.
\end{itemize}

It can be seen that $p(\bm{\theta}| \overline{\bm{y}}, \mathcal{M})$ describes the posterior distribution of the model parameter set $\bm{\theta}$ conditional on measurements $\overline{\bm{y}}$ under model $\mathcal{M}$. The likelihood term gives the probability of observing $\overline{\bm{y}}$ given parameters $\bm{\theta}$. Finally, the denominator in the right hand side is known as the evidence, or marginal likelihood, and gives the likelihood of obtaining the measurements conditional on the model $\mathcal{M}$. Obtaining the evidence is necessary for performing Bayesian model selection. In most practical applications this integral is high-dimensional (see e.g. \cite{Lye2021}) and computationally intractable. Furthermore, the conventional Markov Chain Monte Carlo (MCMC) methods \citep{Metropolis1953,Hastings1970} -- typically used in Bayesian inference -- are primarily geared towards estimating the posterior, and do not compute the evidence. The nested sampling method, implemented in the \textit{Dynesty} Python package \citep{Speagle2019} is utilized to overcome this limitation. In this approach the posterior is separated into nested slices of increasing likelihood. Weighted samples are generated from each slice and subsequently recombined to yield the posterior and evidence. Nested Sampling can deal effectively with moderate to high-dimensional problems (e.g. in problems with up to $100$ parameters) and multi-modal posteriors. The reader is referred to  \citep{Skilling2006} and \citep{Speagle2019} for detailed information on the nested sampling method. 

\subsection{Bayesian model selection}
\label{section:bayesian_model_selection}
The posterior distribution of the parameters for a given set of data is conditional on the model $\mathcal{M}$. Often, multiple models can be defined a priori to describe the observed behaviour. To select the most plausible model, a pool of models $\bm{\mathcal{M}}$ is defined and inference is performed conditional on each model $\mathcal{M}_i$. The Bayes rule can then be applied to select the most likely model based on the evidence. In \cite{Hoeting1999}, the following equation is provided for performing Bayesian model selection:

\begin{equation}
    \label{eq:bayes_model_select}
    p(\mathcal{M}_i|\overline{\bm{y}}) = \frac{p(\overline{\bm{y}}|\mathcal{M}_i) \cdot p(\mathcal{M}_i)}{\sum_{i=1}^{K} p(\overline{\bm{y}}|\mathcal{M}_i) \cdot p(\mathcal{M}_i)},
\end{equation}

\noindent
where:

\begin{itemize}
    \item $p(\mathcal{M}_i|\overline{\bm{y}})$ is the posterior probability of model $i$;
    \item $p(\overline{\bm{y}}|\mathcal{M}_i)$ is the evidence under model $\mathcal{M}_i$;
    \item $p(\mathcal{M}_i)$ is the prior probability of model $i$.
\end{itemize}

Given a pool of models, selecting the model that best fits the data can straightforwardly be achieved by selecting the model that minimizes a particular error metric between measurements and model outputs. However, simply choosing the model that best fits the data could potentially lead to overfitting: more complicated models would tend to fit the data best, making them the most likely in this approach even if the added complexity provides a negligible benefit. An advantage of Bayesian model selection is that it automatically enforces model parsimony, also known as Occam's razor as discussed in \cite{MacKay2003} and \cite{Beck2004}, penalizing overly complex models. It should be emphasized that a high posterior model probability does not necessarily indicate that a particular model provides a good fit with the data, since the model probabilities are conditioned on the pool of candidate models $\bm{\mathcal{M}}$. Therefore, a high posterior model probability can only be interpreted as a particular model being more likely, relative to the other models that are considered. To aid the interpretation of the results, the relative plausibility of two models $\mathcal{M}_1$ and $\mathcal{M}_2$ belonging to a class of models $\bm{\mathcal{M}}$ can be expressed in terms of the Bayes factor:

\begin{equation}
    \label{eq:bayes_factor}
    R = \frac{p(\mathcal{M}_1|\overline{\bm{y}})}{p(\mathcal{M}_2|\overline{\bm{y}})} \cdot \frac{p(\mathcal{M}_2)}{p(\mathcal{M}_1)}
\end{equation}

An advantage of using the Bayes factor over the posterior model probabilities for model selection is that it can be readily interpreted to indicate the support of one model over another, and thus offers a practical means of comparing different models. The interpretation of \cite{Jeffreys2003} is used in this work, given in the table below.

\begin{table}[htb!]
\centering
\caption{Interpretation of the Bayes factor from \cite{Jeffreys2003}.}
\label{tab:K_interpretation}
\begin{tabular}{@{}ll@{}}
\toprule
R                               & Strength of evidence    \\ \midrule
\textless $10^0$ & Negative                \\
$10^0$ to $10^{1/2}$ & Barely worth mentioning \\
$10^{1/2}$ to $10^1$                  & Substantial             \\
$10^1$ to $10^{3/2}$                  & Strong                  \\
$10^{3/2}$ to $10^2$                  & Very strong             \\
\textgreater $10^2$               & Decisive                \\ \bottomrule
\end{tabular}
\end{table}

\subsection{Posterior predictive distribution}
Bayesian system identification can be used to obtain point estimates and posterior distributions of uncertain parameters using physical models and measurement data. However, directly using the point estimates of the inferred parameters to make predictions would result in underestimation of the uncertainty and overly confident predictions. This is due to the fact that using point estimates to make predictions disregards the uncertainty in the inferred parameters resulting from lack of data. In contrast, the posterior predictive is a distribution of possible future observations conditioned on past observations taking into account the combined uncertainty from all sources (e.g. modeling and measurement error and parameter uncertainty). The posterior predictive can be obtained as \citep{Gelman2013}: 

\begin{equation}
    \label{eq:posterior_predictive}
    p(\bm{y}^*|\overline{\bm{y}}) = \int_{\Theta} p(\bm{y}^*|\bm{\theta}) \cdot p(\bm{\theta}|\overline{\bm{y}}) \cdot \text{d}\bm{\theta},
\end{equation}

\noindent
where $\bm{y}^*$ is a vector of possible future observations.

\subsection{Data generating process}
\label{section:probabilistic_models}

In order to perform system identification, the likelihood function is formulated based on the combination of a probabilistic model and a deterministic physical model. This coupled probabilistic-physical model is used to represent the process that is assumed to have generated the measurements, referred to as the \textit{data generating process}. Details on the deterministic physical model are provided in Section \ref{section:physical_model}. The probabilistic model is used to represent the uncertainties that are inherent when using a model to describe a physical system. The following sources of uncertainty are considered:\\

\begin{itemize}
    \item Measurement uncertainty
    \item Physical model uncertainty
\end{itemize}

Measurement uncertainty refers to the error between the measured response quantities and the true system response, caused by the combined influence of sensor errors and environmental noise \citep{Kennedy2001}. Modeling uncertainty can contain several components and refers to the error between reality and the models used to represent it. These errors arise e.g. due to simplifications in the physical model and numerical approximations.

In this paper, we consider data generating processes based on a multiplicative and additive model prediction error, which will be explained in the following subsections. In the following expressions, Greek letters are used to represent random variables, while bold lower and upper-case letters denote vectors and matrices respectively.  

\subsubsection{Multiplicative model}
\label{section:multiplicative_model}
The data generating processes described by \autoref{eq:multiplicative_model} is obtained by considering the discrepancies between the deterministic model output and the real system response, a process referred to as \textit{stochastic embedding} in \cite{Beck2010}. In this model of the data generating process, a multiplicative prediction error is considered:

\begin{equation}
    \label{eq:multiplicative_model}
    \bm{\chi}(\bm{\theta}) = \bm{Y}(\bm{\theta}_{\mathrm{s}}) \bm{\eta}_{\mathrm{m}}(\bm{\theta}_{\mathrm{c}}) + \bm{\epsilon}(\bm{\theta}_{\mathrm{c}}),
\end{equation}

\noindent
where:

\begin{itemize}
    \item $\bm{\chi}$ is a vector of predictions obtained from the coupled physical-probabilistic model of the data generating process;
    \item $\bm{Y}$ is a diagonal matrix of physical model predictions obtained as $\bm{Y} = \mathrm{diag}(\bm{y})$, with $\bm{y}$ denoting the corresponding vector of predictions;
    \item $\bm{\eta}_{\mathrm{m}}$ is a vector of multiplicative physical model error factors;
    \item $\bm{\epsilon}$ is a vector of measurement error random variables;
    \item $\bm{\theta}_\mathrm{s}$ is a vector of physical model parameters to be estimated;
    \item $\bm{\theta}_\mathrm{c}$ is a vector of probabilistic model parameters to be estimated;
    \item $\bm{\theta} = \{\bm{\theta}_{\mathrm{s}}, \bm{\theta}_{\mathrm{c}}\}$ is the set of combined physical and probabilistic model parameters to be estimated.
\end{itemize}

In this model formulation, the error in the physical model prediction is assumed to scale with the magnitude of the model output. This assumption is prevalent in the structural reliability literature \citep{Cervenka2018,Sykorka2018}. The physical model predictions are multiplied by a factor $\bm{\eta}_{\mathrm{m}}$, expressing the discrepancy between model prediction and reality. A correlated Multivariate Normal distribution with a mean of $1.0$ and covariance matrix $\bm{\Sigma}_{\mathrm{\eta}}$ is assumed for $\bm{\eta}_{\mathrm{m}}$ as shown in \autoref{eq:k_model_dist}: 

\begin{equation}
    \label{eq:k_model_dist}
    \bm{\eta}_{\mathrm{m}}(\bm{\theta}_{\mathrm{c}}) \sim \mathcal{N}(1.0, \bm{\Sigma}_{\eta}(\bm{\theta}_{\mathrm{c}}))
\end{equation}

\begin{figure}[htb!]
    \centering
    \FIG{\includegraphics[width=0.60\textwidth]{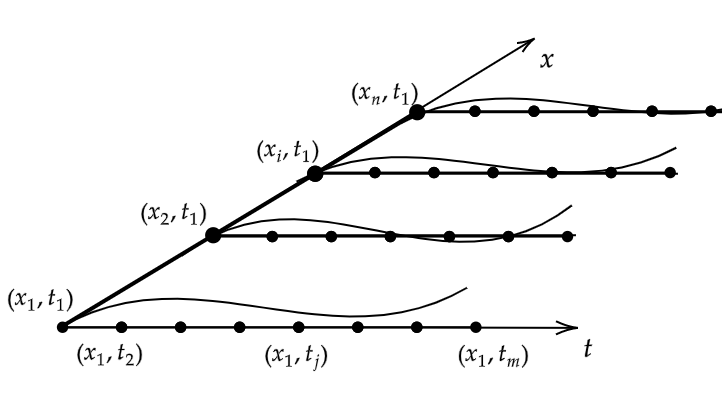}}
    \caption{Illustration of space and time coordinate system. Influence lines along the time axis $t$ are obtained for each sensor position $x$.}
    \label{fig:spacetime_coords}
\end{figure}

The assumption of a Gaussian distribution for $\bm{\eta}_{\mathrm{m}}$ is made primarily for simplicity and computational convenience. The impact of this assumption is deemed to be outside the scope of this work and is not further examined. The measurement error is taken as independent, identically distributed (i.i.d.) Gaussian random variables, distributed as $\bm{\epsilon} \sim \mathcal{N}(0, \sigma_{\mathrm{\epsilon}})$. The assumption of Gaussian white noise for the measurement error is prevalent in the literature and is commonly used in Bayesian system identification for structures (see Section \ref{section:literature}), stemming from the fact that measurement noise can be considered as a sum of a large number of independent random variables. Modeling the measurement error as i.i.d. realizations from a Normal distribution is therefore justified by the central limit theorem. Utilizing the affine transformation property of the Multivariate Normal distribution we obtain the following model for the data generating process:


\begin{equation}
    \label{eq:distribution_multiplicative}
    \bm{\chi}_{\mathrm{m}} \sim \mathcal{N}(\: \bm{y}(\bm{\theta}_\mathrm{s}), \: \bm{Y} \bm{\Sigma}_{\eta} \bm{Y}^T + \sigma_{\mathrm{\epsilon}}^2 \bm{I} \: ),
\end{equation}

with $\bm{I}$ being the identity matrix. The residuals between measurements and model predictions are considered as a random field, with the position of each observation defined by a spatial coordinate (representing the location of a sensor) and a temporal coordinate, denoted as $x_i$ and $t_i$ respectively. The position of an observation $\overline{\bm{y}}$ is described by a two-dimensional vector $\bm{x}_i = (x_i, t_i)$, and the random field is represented as a (not necessarily regular) grid of points, as shown in \autoref{fig:spacetime_coords} with $n$ denoting the total number of sensors and $m$ denoting the number of observations per sensor over time. 

The correlation in the model prediction error between two points $\bm{x}_i = (x_i,t_i)$ and $\bm{x}_j = (x_j, t_j)$ is obtained as the product of the spatial and temporal correlation, described in terms of the respective kernel functions: 

\begin{equation}
    \label{eq:correaltion_function}
    \rho_{i,j} = k_x(x_i, x_j ;\bm{\theta}_\mathrm{c}) \cdot k_t(t_i, t_j ; \bm{\theta}_\mathrm{c}),
\end{equation}

\noindent
where $k_x(x_i, x_j ;\bm{\theta}_\mathrm{c})$ and $k_t(t_i, t_j ;\bm{\theta}_\mathrm{c})$ are parametrized by the set of parameters of the probabilistic model $\bm{\theta}_\mathrm{c}$. The standard deviation of the model prediction error at a point $i$ is obtained as $\sigma_i = C_{\mathrm{v}} \cdot y_{i}$, where $C_{\mathrm{v}}$ denotes the coefficient of variation (COV) of the model prediction error. Calculating the covariance for every pair of points yields a symmetric positive semi-definite covariance matrix $\bm{\Sigma}_{\mathrm{\eta}}$ that describes the covariance of the physical model prediction error for every point in the random field:

\begin{equation}
    \label{eq:cov_mx}
    \bm{\Sigma}_{\mathrm{p}} = \bm{Y} \bm{\Sigma}_{\mathrm{\eta}} \bm{Y}^T = \begin{bmatrix} \sigma_{1}^2 & \hdots & \sigma_1 \cdot \sigma_N \cdot \rho_{1,N} \\ \vdots & \ddots & \vdots  \\ \sigma_N \cdot \sigma_1 \cdot \rho_{N,1} & \hdots & \sigma_N^2  \end{bmatrix},
\end{equation}

\noindent
with $N = n \cdot m$.

\subsubsection{Additive model}
\label{section:additive_model}
A coupled probabilistic-physical model of the data generating process based on a correlated, additive model prediction error is also considered (\autoref{eq:additive_model}). In this case the model prediction error is described by an additive term, modeled as a multivariate normal distribution with zero mean and covariance $\bm{\Sigma}_{\mathrm{\eta}} (\bm{\theta}_{\mathrm{c}})$. Similarly to the multiplicative model, the measurement error is represented by a vector of i.i.d. normal random variables, with a mean of zero and standard deviation $\sigma_{\mathrm{\epsilon}}$. 

\begin{equation}
    \label{eq:additive_model}
    \bm{\chi}_{\mathrm{a}}(\bm{\theta}) = \bm{y}(\bm{\theta}_{\mathrm{s}}) + \bm{\eta_{\mathrm{a}}}(\bm{\theta}_{\mathrm{c}}) + \bm{\epsilon}(\bm{\theta}_{\mathrm{c}}),
\end{equation}

\noindent
where $\bm{\eta}_{\mathrm{a}}$ denotes the vector of additive model prediction error.

\begin{equation}
    \label{eq:a_model_dist}
    \bm{\eta}_{\mathrm{a}}(\theta_{\mathrm{c}}) \sim \mathcal{N}( \: 0.0, \: \bm{\Sigma}_{\mathrm{\eta}}(\bm{\theta}_{\mathrm{c}}) + \sigma^2_{\mathrm{\epsilon}} \bm{I} \: )
\end{equation}
\section{Efficient log-likelihood evaluation}
\label{section:efficient_likelihood}

\noindent
The assumption of a multiplicative or additive physical model uncertainty factor described by a Gaussian distribution leads to a multi-variate Gaussian likelihood description. For a given covariance matrix $\bm{\Sigma}$, and omitting the dependence on the parameter vector $\bm{\theta}$ from the right hand side of the equation for brevity, the multivariate normal log-likelihood function can be expressed as:

\begin{equation}
    \label{eq:loglike}
    \mathcal{L}(\bm{\theta}) = -\frac{1}{2} \cdot \left[ \log {|\bm{\Sigma}|} + (\overline{\bm{y}} - \bm{y})^T \bm{\Sigma}^{-1}(\overline{\bm{y}} - \bm{y}) + N \cdot \log{(2 \cdot \pi)} \right],
\end{equation}

The evaluation of the Multivariate Gaussian log-likelihood (\autoref{eq:loglike}) requires calculating the determinant (${|.|}$) and inverse (${(.)^{-1}}$) of the covariance matrix $\bm{\Sigma}$. These operations typically have $O(N^3)$ time complexity and $O(N^2)$ memory requirements for factorizing and storing the covariance matrix respectively, making the direct evaluation of the log-likelihood infeasible for more than a few thousand points. 

To address this issue, we present an approach for efficient log-likelihood evaluation under the ${\it multiplicative}$ model uncertainty with additive Gaussian noise described in Section \ref{section:multiplicative_model}. For the case of ${\it additive}$ model uncertainty, described in Section \ref{section:additive_model} we utilize an existing approach from the literature. A comparison of the average wall clock time required for log-likelihood evaluation as a function of the size of a 2-dimensional grid of measurements against a naive implementation using the full covariance matrix, for both the additive and multiplicative cases, can be found in \cite{Koune2021}. A Python implementation of both methods is available at \url{https://github.com/TNO/tripy}.

\subsection{Efficient log-likelihood evaluation for combined spatial and temporal correlation and multiplicative model prediction uncertainty}
\label{section:efficient_loglikelihood_multiplicative}
To reduce the computational complexity for evaluating the log-likelihood under the ${\it multiplicative}$ model uncertainty, we propose an approach that utilizes the tridiagonal inverse form of the correlation matrix that can be obtained from the Exponential kernel, as well as the Kronecker structure of the separable space and time covariance matrix.

In the following it is assumed that the correlation is exponential in time. No assumptions are made regarding the structure of the correlation in space or the number of spatial dimensions. The $i,j^{\mathrm{th}}$ element of the temporal covariance matrix $\bm{\Sigma}_\mathrm{t}$ is obtained as:

\begin{equation}
    \label{eq:exponential}
    \bm{\Sigma}_{\mathrm{t}}^{i,j} = C_{\mathrm{v},i} \cdot C_{\mathrm{v},j} \cdot \exp \left( \frac{\|t_i - t_j\|}{l_{\mathrm{corr}}} \right),
\end{equation}

\noindent
where $l_{\mathrm{corr}}$ is the correlation length and $C_{\mathrm{v}}$ is the coefficient of variation described in Section \ref{section:multiplicative_model}. It is shown in \cite{Pasquier2020} that the inverse of the covariance matrix for this kernel function has a symmetric tridiagonal form:

\begin{equation}
\bm{\Sigma}_{\mathrm{t}}^{-1} = \begin{bmatrix} d_1 & c_1 &  &  & \ &  \\ c_1 & d_2 & c_2 &  &  &  \\  & c_2 & d_3 & c_3 &  &  \\ &  & \ddots & \ddots & \ddots \end{bmatrix}
\label{eq:C_inv_symmetric}
\end{equation}

\noindent
Following \cite{Cheong2016}, the diagonal vectors of diagonal and off-diagonal terms in \autoref{eq:C_inv_symmetric} can be obtained analytically, eliminating the need to form the full correlation and covariance matrices which is often computationally intensive due to the amount of memory and operations required. For a given vector of observations with coordinates $\bm{t} = \{t_1, t_2, ..., t_m \}$ denoting $\Delta t_i = |t_i - t_{i-1}|$ for $i \in [m]$ yields \autoref{eq:corr_alpha}.

\begin{equation}
    \label{eq:corr_alpha}
    a_i = e^{- \lambda \cdot \Delta t_i},
\end{equation}

\noindent
where $\lambda$ is the inverse of the correlation length $l_{\mathrm{corr}}$ and $a_i$ is the correlation between points $i$ and $i-1$. The diagonal and off-diagonal elements of the inverse correlation matrix $\bm{\Sigma}_\mathrm{t}^{-1}$ can then be obtained analytically, eliminating the need for direct inversion of $\bm{\Sigma}_{\mathrm{t}}$ and reducing computational complexity and memory requirements: 

\begin{gather}
    d_{1} = \frac{1}{C_{\mathrm{v},1}^2} \cdot \frac{1}{1 - a_2^2}, \\
    d_{m} = \frac{1}{C_{\mathrm{v},m}^2} \cdot \frac{1}{1 - a_{m}^2}, \\
    d_{ii} = \frac{1}{C_{\mathrm{v},i}^2} \cdot \left( \frac{1}{1 - a_i^2} + \frac{1}{1 - a^2_{i+1}} - 1 \right), \\
    c_{ii-1} = - \frac{1}{C_{\mathrm{v},i} \cdot C_{\mathrm{v},i+1}} \cdot \frac{a_i}{1-a_i^2}
\end{gather}

Furthermore, we define the combined space and time covariance which can be obtained as the Kronecker product of the temporal correlation matrix and the spatial correlation matrix, $\bm{\Sigma}_{\mathrm{\eta}} = \bm{\Sigma}_t \otimes \bm{\Sigma}_x$. Using the properties of the Kronecker product, it can be shown that the resulting inverse matrix $\bm{\Sigma}_{\mathrm{\eta}}^{-1}$ has a symmetric block tridiagonal form:


\begin{equation}
\bm{\Sigma}_{\mathrm{\eta}}^{-1} = \begin{bmatrix} \bm{D}_1 & \bm{C}_1 &  &  & \ &  \\ \bm{C}_1 & \bm{D}_2 & \bm{C}_2 &  &  &  \\  & \bm{C}_2 & \bm{D}_3 & \bm{C}_3 &  &  \\ &  & \ddots & \ddots & \ddots \end{bmatrix}
\end{equation}

We consider the covariance matrix for the data generating process defined in \autoref{eq:distribution_multiplicative}. Expressing the physical model uncertainty covariance matrix $\bm{\Sigma}_{\mathrm{p}}$ in terms of the combined space and time covariance $\bm{\Sigma}_{\mathrm{\eta}}$ yields:


\begin{equation}
    \label{eq:cov_phys_expanded}
    \bm{\Sigma}_{\mathrm{p}} = \bm{Y} \bm{\Sigma}_{\mathrm{\eta}} \bm{Y} ^T
\end{equation}

Then $\bm{\Sigma}_{\mathrm{p}}^{-1}$ will also be block tridiagonal. However, including additive noise such that $\bm{\Sigma}_{\mathrm{p}} = \bm{Y} \bm{\Sigma}_{\mathrm{\eta}} \bm{Y}^T + \sigma^2_{\mathrm{\epsilon}} \bm{I}$ leads to a dense inverse matrix. To efficiently evaluate the likelihood, we aim to calculate the terms $(\overline{\bm{y}} - \bm{y})^T \bm{\Sigma}_{\mathrm{p}}^{-1} (\overline{\bm{y}} - \bm{y})$ and $|\bm{\Sigma}_{\mathrm{p}}|$ in \autoref{eq:loglike} without explicitly forming the corresponding matrices or directly inverting the covariance matrix, while taking advantage of the properties described previously to reduce the complexity. Algebraic manipulation of the product $(\overline{\bm{y}} - \bm{y})^T \bm{\Sigma}_{\mathrm{p}}^{-1} (\overline{\bm{y}} - \bm{y})$ is performed in order to obtain an expression that can be evaluated efficiently by taking advantage of the Kronecker structure and block symmetric tridiagonal inverse of the covariance matrix. We apply the Woodbury matrix identity given below:

\begin{equation}
    \label{eq:woodbury}
    (\bm{A}^{-1} + \bm{B}\bm{C}^{-1}\bm{B}^T)^{-1} = \bm{A} - \bm{A}\bm{B}(\bm{C} + \bm{B}^T \bm{A} \bm{B})^{-1} (\bm{A}\bm{B})^T
\end{equation}

\noindent
Substituting $\bm{A} \to \bm{\Sigma}_{\mathrm{\epsilon}}^{-1}$, $\bm{B} \to \bm{Y}$ and $\bm{C}^{-1} \to \bm{\Sigma}_{\mathrm{\eta}}$, yields:

\begin{equation}
    \label{eq:rearanged}
    \bm{\Sigma}^{-1}_{\mathrm{p}} = \bm{\Sigma}_{\mathrm{\epsilon}}^{-1} - (\bm{\Sigma}_{\mathrm{\epsilon}}^{-1} \bm{Y})(\bm{\Sigma}_{\mathrm{\eta}}^{-1} + \bm{Y}^{T}\bm{\Sigma}_{\mathrm{\epsilon}}^{-1}\bm{Y})^{-1} (\bm{\Sigma}_{\mathrm{\epsilon}}^{-1}\bm{Y})^T,
\end{equation}

\noindent
Applying the left and right vector multiplication by $\bm{y}$, the second term in the r.h.s. of \autoref{eq:loglike} becomes:

\begin{equation}
    \label{eq:multiplied}
    \bm{y}^{T} \bm{\Sigma}_{\mathrm{p}}^{-1} \bm{y} = \bm{y}^T \bm{\Sigma}_{\mathrm{\epsilon}}^{-1}\bm{y} - \bm{y}^T (\bm{\Sigma}_{\mathrm{\epsilon}}^{-1} \bm{Y})(\bm{\Sigma}_{\mathrm{\eta}}^{-1} + \bm{Y}^T \bm{\Sigma}_{\mathrm{\epsilon}}^{-1} \bm{Y})^{-1}(\bm{\Sigma}_{\mathrm{\epsilon}}^{-1}\bm{Y})^T \bm{y}
\end{equation}

\noindent
In the previous expression, the term $\bm{y}^T \bm{\Sigma}_{\mathrm{\epsilon}}^{-1}\bm{y}$ can be efficiently evaluated as the product of vectors and diagonal matrices. Similarly, the term $\bm{y}^T (\bm{\Sigma}_{\mathrm{\epsilon}}^{-1} \bm{Y})$ can be directly computed and yields a vector. We consider the following term from the r.h.s. of \autoref{eq:multiplied}:

\begin{equation}
    \label{eq:tridiag_terms}
    (\bm{\Sigma}_{\mathrm{\eta}}^{-1} + \bm{Y}^T \bm{\Sigma}_{\mathrm{\epsilon}}^{-1} \bm{Y})^{-1}(\bm{\Sigma}_{\mathrm{\epsilon}}^{-1}\bm{Y})^T \bm{y} = \bm{X} 
\end{equation}


We note that the term $\bm{\Sigma}_{\mathrm{\eta}}^{-1} + \bm{Y}^T \bm{\Sigma}_{\mathrm{\epsilon}}^{-1} \bm{Y}$ is the sum of a symmetric block tridiagonal matrix $\bm{\Sigma}_{\mathrm{\eta}}^{-1}$ and the diagonal matrix $\bm{Y}^T \bm{\Sigma}_{\mathrm{\epsilon}}^{-1} \bm{Y}$. We can therefore take advantage of efficient algorithms for Cholesky factorization of symmetric block tridiagonal matrices and for solving linear systems using this factorization to compute $\bm{X}$. Furthermore, the Cholesky factors obtained previously are also used to reduce the computational cost of evaluating the determinant $|\bm{\Sigma}| = |\bm{\Sigma}_{\mathrm{\epsilon}} + \bm{Y} \bm{\Sigma}_{\mathrm{\eta}} \bm{Y}^T|$. Applying the determinant lemma for $\bm{\Sigma}_{\mathrm{p}}$ yields \autoref{eq:determinant_simple}.


\begin{equation}
    \label{eq:determinant_simple}
    |\bm{\Sigma}_{\mathrm{\epsilon}} + \bm{Y} \bm{\Sigma}_{\mathrm{\eta}} \bm{Y}^T| = |\bm{\Sigma}_{\mathrm{\eta}}^{-1} + \bm{Y}^T \bm{\Sigma}_{\mathrm{\epsilon}}^{-1}\bm{Y}| \cdot |\bm{\Sigma}_{\mathrm{\eta}}| \cdot |\bm{\Sigma}_{\mathrm{\epsilon}}|
\end{equation}

\noindent
The determinant $|\bm{\Sigma}_{\mathrm{\eta}}|$ can be calculated efficiently by utilizing the properties of the Kronecker product, given that $|\bm{\Sigma}_{\mathrm{\eta}}| = |\bm{\Sigma}_t \otimes \bm{\Sigma}_x|$. Furthermore, $\bm{\Sigma}_{\mathrm{\epsilon}}$ is a diagonal matrix meaning that the determinant can be trivially obtained. Finally, we have previously calculated the Cholesky factorization of the term $\bm{\Sigma}_{\mathrm{\eta}}^{-1} + \bm{Y}^T \bm{\Sigma}_{\mathrm{\epsilon}}^{-1}\bm{Y}$. Using the fact that the determinant of a block triangular matrix is the product of the determinants of its diagonal blocks and the properties of the determinant, the first expression in the r.h.s. of \autoref{eq:determinant_simple} can be computed with:

\begin{equation}
    |\bm{\Sigma}_{\mathrm{\eta}}^{-1} + \bm{Y}^T \bm{\Sigma}_{\mathrm{\epsilon}}^{-1}\bm{Y}| = |\bm{LL}^T| = |\bm{L}|\cdot |\bm{L}^T|= |\bm{L}|^2,
\end{equation}

\noindent
where the matrix $\bm{L}$ is the lower triangular Cholesky factor of $\bm{\Sigma}_{\mathrm{\eta}}^{-1} + \bm{Y}^T \bm{\Sigma}_{\mathrm{\epsilon}}^{-1}\bm{Y}$. Since each block $\bm{L}_{ii}$ is also triangular, the evaluation of the determinant has been reduced to evaluation of the determinant of each triangular block $\bm{L}_{ii}$, which is equal to the product of its diagonal elements. 

Using the above calculation procedure, an efficient solution can also be obtained for the case of only temporal correlation, where $\bm{\Sigma}_t^{-1}$ has the symmetric tridiagonal form given in \autoref{eq:C_inv_symmetric}. The term $\bm{\Sigma}_t^{-1} + \bm{Y}^T \bm{\Sigma}_{\mathrm{\epsilon}}^{-1} \bm{Y}$ will be the sum of a symmetric tridiagonal and a diagonal matrix. From a computational viewpoint, this property is advantageous as it allows for a solution to the system of equations (\autoref{eq:tridiag_terms}) with $O(N)$ operations using the Thomas algorithm \citep{Quarteroni2007}. Alternatively, for improved efficiency and numerical stability a Cholesky decomposition can be applied to solve the linear system and calculate the determinants of the symmetric tridiagonal terms in \autoref{eq:determinant_simple}. 

\subsection{Efficient log-likelihood evaluation for combined spatial and temporal correlation and additive model prediction uncertainty}
To reduce the computational complexity of the log-likelihood evaluation in the case of ${\it additive}$ model prediction uncertainty and combined spatial and temporal correlation, we use an approach that utilizes the properties of the Kronecker product and the eigendecomposition of the separable covariance matrix. For a detailed description of this approach, the reader is referred to \cite{Stegle2011}. 

\section{Description of the IJssel bridge case study}
\label{section:case_study}

\subsection{Description of the structure}
\begin{figure}[htb!]
    \centering
    \FIG{\includegraphics[width=1.00\textwidth]{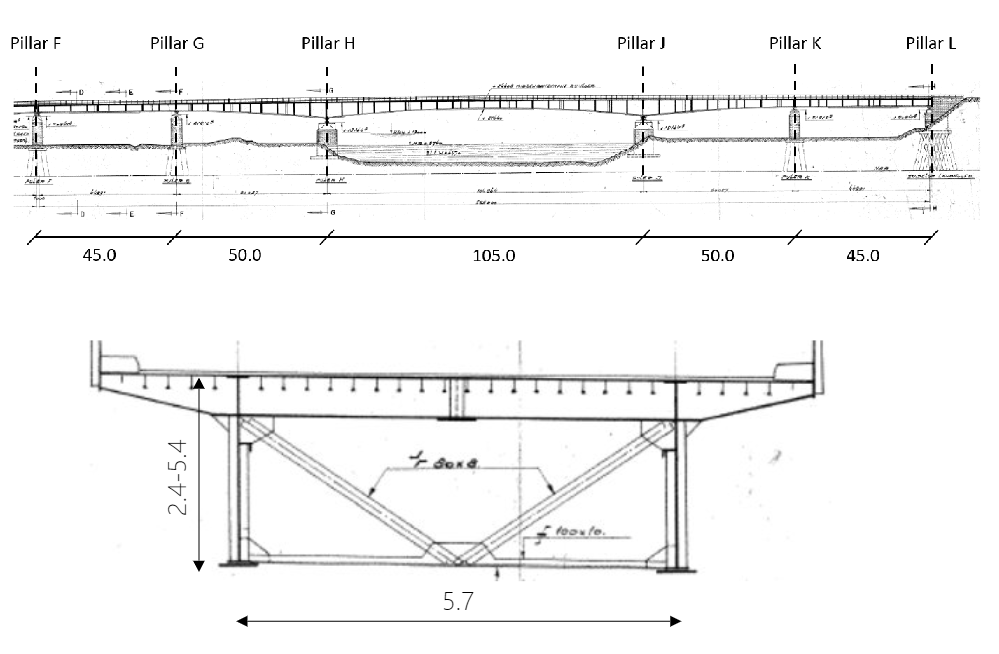}}
    \caption{Elevation view of the IJsselbridge (top), and typical cross-section including K-brace (bottom), with lengths shown in meters (from Rijkswaterstaat).}
    \label{fig:elevation_and_cross_section}
\end{figure}

The IJssel bridge is a twin-girder steel plate road bridge that carries traffic over the river IJssel in the direction of Westervoort. It consists of an approach bridge and a main bridge, of which the latter is of interest in this case. The main bridge has a total length of 295 m and five spans with lengths of 45 m, 50 m, 105 m, 50 m and 45 m. In total the bridge has 12 supports. An elevation view of the structure is shown in \autoref{fig:elevation_and_cross_section}. The supports at pillar H are hinges, while the rest are roller bearings in the longitudinal directions. The roller bearings at pillars G and K can resist uplift forces. The deck structure of the bridge is composed of two steel girders with variable height, ranging from 2.4 to 5.3 m, and cross-beams with a spacing of approximately 1.8 m. The main girders and cross-beams support the steel deck plate. The deck plate has a thickness of 10 or 12 mm and $160 \times 8$ mm longitudinal bulb stiffeners. The cross beams are placed with a center-to-center distance of $1.75$ to $1.80$ m and are composed of a $500 \times 10$ mm web with a $250 \times 12$ mm welded flange. The cross beams are tapered in the parts that extend beyond the main girders and the beam height is reduced to $200$ mm at the beam ends. The two main girders are coupled with K-braces located below every second or third cross beam, with a distance of $5.4$ meters on average.

\subsection{Description of the physical model}
\label{section:physical_model}
\begin{figure}[htb!]
     \centering
     \begin{subfigure}[b]{1.0\textwidth}
         \centering
        \includegraphics[width=0.95\textwidth]{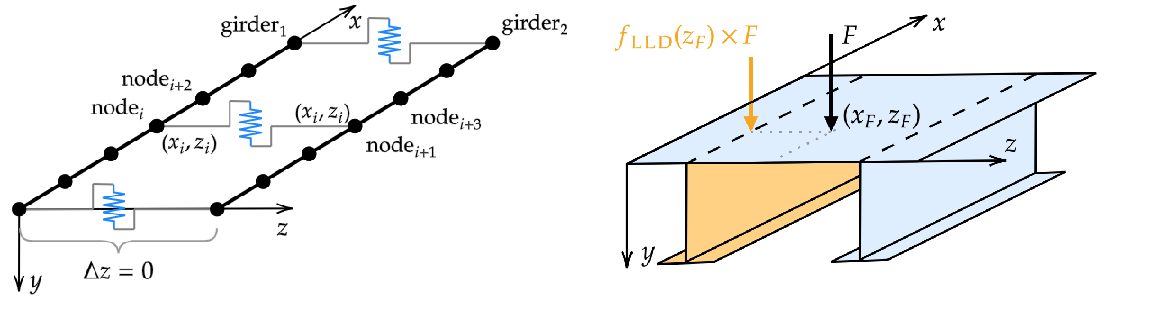}
         \label{fig:model_and_LLF_1}
     \end{subfigure}
     \hfill
     
     \begin{subfigure}[b]{1.0\textwidth}
         \centering
         \includegraphics[width=0.95\textwidth]{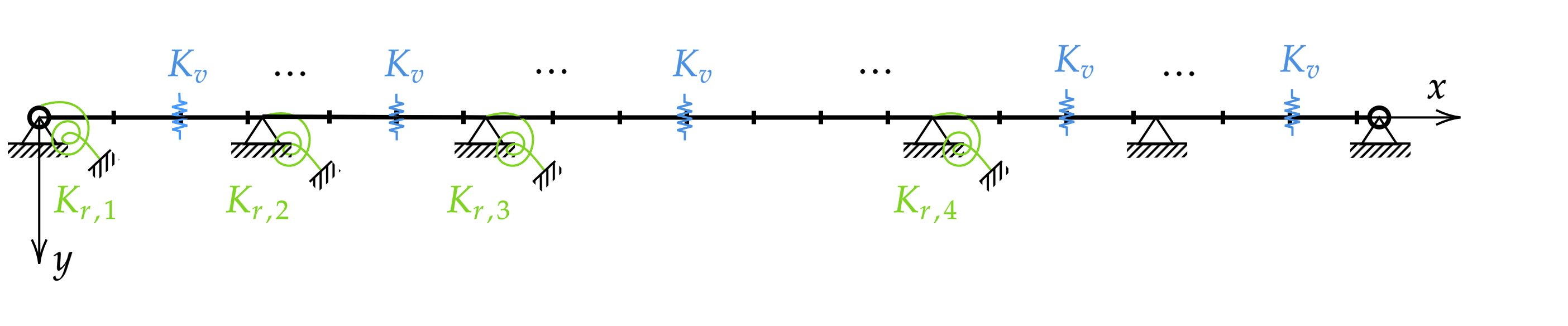}
         \label{fig:model_and_LLF_2}
     \end{subfigure}

        \caption{Illustration of the IJsselbridge FE model (left), lateral load function (right), and parametrization of the FE model (bottom).}
        \label{fig:model_and_LLF}
\end{figure}

A two-dimensional twin-girder finite element (FE) model based on Euler-Bernoulli beam elements is used to model the IJsselbridge, shown in \autoref{fig:model_and_LLF}. Each element has four degrees of freedom (DOFs): two translations and two rotations. The variable geometrical properties of the steel girders along the longitudinal axis are taken into account by varying the structural properties of the individual beam elements, where each element has a prismatic cross section. In addition to the main girder, half the width of the deck plate and the corresponding longitudinal stiffeners are also considered in the calculation of the structural properties for each cross section along the $x$ axis. The maximum beam element length can be specified in order to approximate the variable geometry of the main girders along the length of the bridge to the required precision. A maximum element length of $2.0$ m was used for all simulations, resulting in a model with $n_{\mathrm{node}} = 193$ nodes and $n_{\mathrm{DOF}} = 386$ DOFs. The coupling between the main girders due to combined stiffness of the deck, crossbeams and K-braces is simulated by vertical translational springs, placed approximately at the positions of the K-braces that connect the two main girders of the IJssel bridge. Six pinned supports are specified for each girder at locations corresponding to pillars F through L. Independent rotational springs are defined at the supports to simulate the friction at the support bearings and to account for the possibility of partial locking. 

During the measurement campaign, the bridge was closed for traffic and only loaded by a heavy weighted truck at the left and right lane. To account for the position of the truck along the transverse direction ($z$-axis) -- which is not included in the FE model -- each load is multiplied by a value of the Lateral Load Function (LLF), as illustrated in \autoref{fig:model_and_LLF} (right). The LLF is taken as a linear function, with slope and intercept coefficients such that: (i) a point load applied directly on a girder does not affect the other girder; and (ii) a load applied at the center of the bridge deck is equally distributed between the left and right girders.

\subsection{Measurements}
\label{section:fugro_measurements}
The data used in this study is obtained from a measurement campaign performed under controlled load tests. A total of 34 strain measurement sensors were placed on the top and bottom flanges of the steel girders, the cross beams, the longitudinal bulb stiffeners and the concrete approach bridge, to measure the response of the structure to traffic loads. In this paper, we use only a subset of sensors that are placed on the center of the bottom flange of the right main girder, since they measure the global response of the bridge. These sensors are denoted with the prefix "H". The exact position of each considered sensor along the length of the bridge and the sensor label are provided in \autoref{tab:sensors_fugro}. It should be noted that the authors were given access to the experimental data and details regarding the sensor network, data acquisition system, and experimental procedure of the measurement campaign, however, the location of the sensors was not chosen specifically for the purposes of the present study. Additional information regarding the measurement campaign can be found in Appendix \ref{appendix:measurements}.

\noindent
\begin{figure}[htb!]
    \centering
    \FIG{\includegraphics[width=1.00\textwidth]{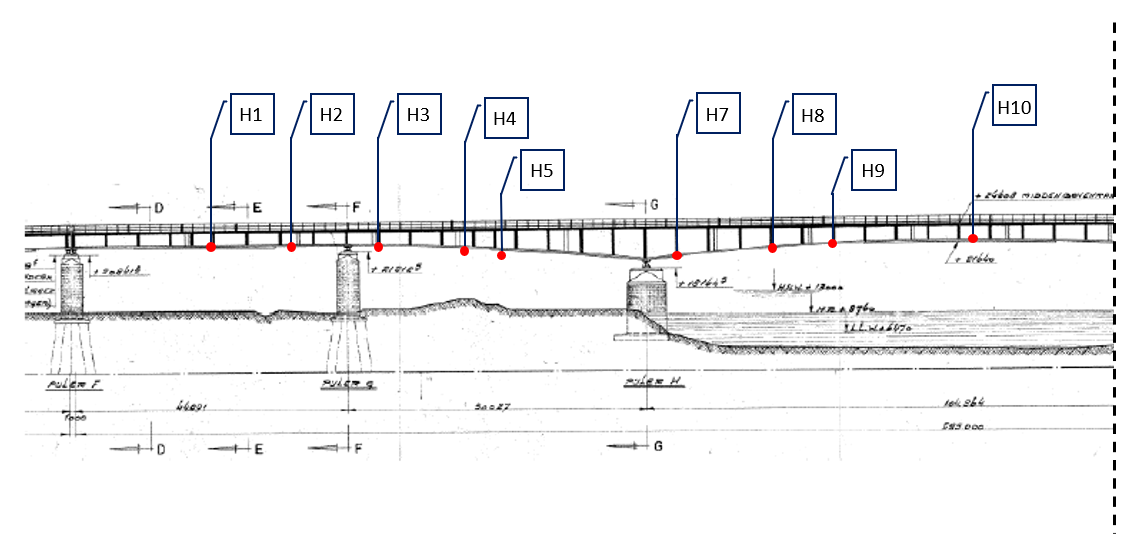}}
    \caption{Approximate location of sensors on the right girder. The prefix "H" is used to denote the sensors on the main structure of the IJsselbridge. Adapted from a Rijkswaterstaat internal report.}
    \label{fig:fugro_sensor_locations}
\end{figure}

\begin{table}[htb!]
\centering
\caption{Names, labels and positions of strain gauges placed on the IJsselbridge main girder. The positions are measured from pillar F (see \autoref{fig:elevation_and_cross_section}).}
\label{tab:sensors_fugro}
\begin{tabular}{@{}cccccccccc@{}}
\toprule
Sensor & H1 & H2 & H3 & H4 & H5 & H7 & H8 & H9 & H10 \\ \midrule
Position [m] & 20.42 & 34.82 & 47.70 & 61.97 & 68.60 & 96.80 & 113.35 & 123.90 & 147.50 \\
\bottomrule
\end{tabular}
\end{table}

\begin{figure}[htb!]
    \centering
    \FIG{\includegraphics[width=0.9\textwidth]{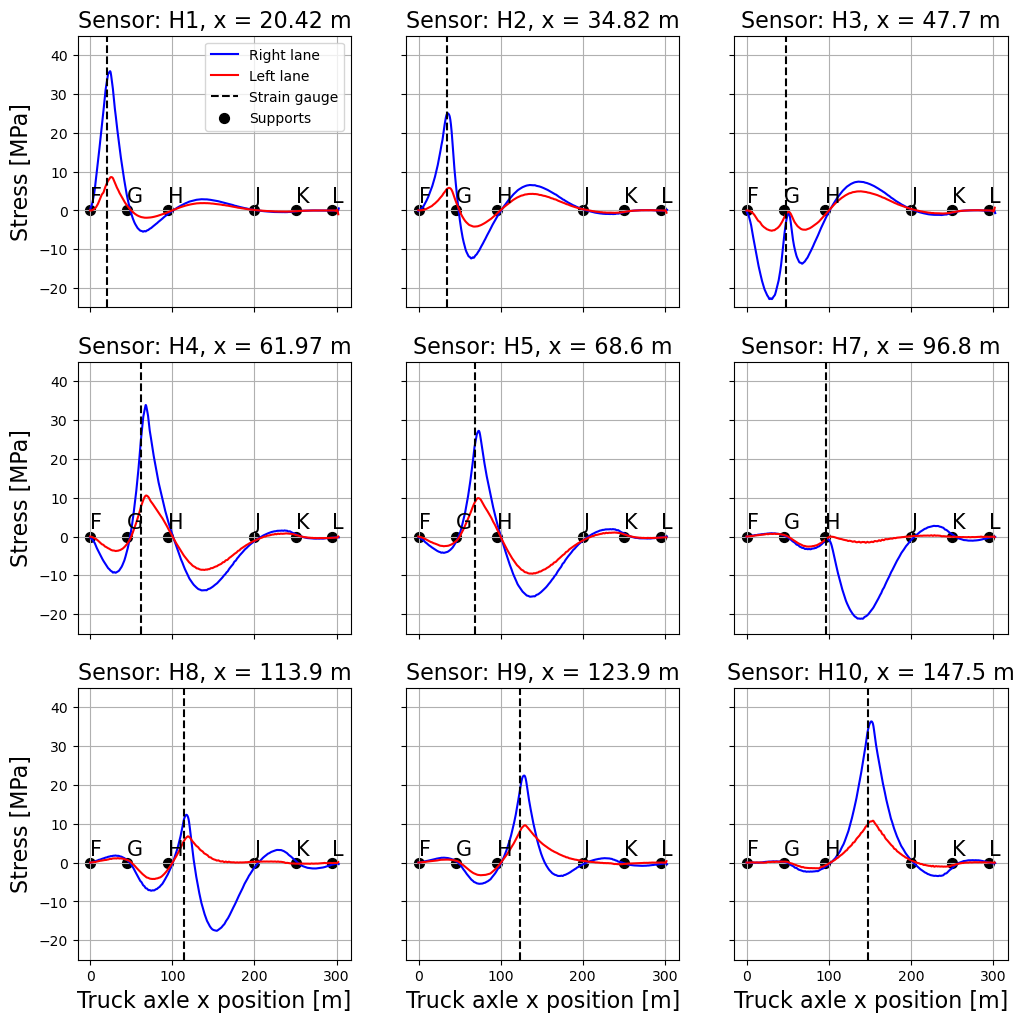}}
    \caption{Stress influence lines obtained from the measurement campaign. The blue and red lines correspond to the measured response for different truck positions in the transverse direction of the bridge.}
    \label{fig:fugro_influence_lines}
\end{figure}

Time series of the strain $\varepsilon$ at a sampling rate of $50$ Hz are obtained from each sensor and postprocessed to yield the corresponding influence lines. These strain influence lines are converted to stress influence lines using Hooke's law $\sigma = E \cdot \varepsilon$, where $\sigma$ denotes the stress and $E$ denotes Young's modulus. The latter is taken as $E = 210$ GPa, as specified in the IJsselbridge design. Each sensor yields two influence lines, one for each lane that was loaded during the measurement campaign. Linear interpolation is performed to obtain the stresses at the locations along the length of the bridge corresponding to the locations of the nodes of the FE model. The processed influence lines are plotted in \autoref{fig:fugro_influence_lines}.

It is noted that significant discrepancies were observed between the measurements and FE model predictions for a number of sensors during a preliminary comparison of model predictions using a model fitted to the measurements with conventional numerical optimization. It was determined, after verifying the validity of the measurements, that this can be attributed to the simplicity of the model, which was not able to capture the structural behaviour at a number of sensor locations. The following list is a summary of the physical simplifications and assumptions that could potentially contribute to the observed discrepancies:

\begin{itemize}
    \item Only limited number of structural elements are explicitly modelled, with elements such as stiffeners, K-braces, the steel deck and cross-beams being only considered implicitly (e.g. by modifying the cross-sectional properties of the elements representing the main girders), or omitted entirely.
    \item The 3D distribution of forces within the elements is neglected, loads are lumped to the closest node and supports are considered as points, potentially overestimating the stresses.
    \item Although likely negligible, the stiffness of the deck and cross-beams between the two girders is only considered implicitly as lumped stiffness in the vertical springs.
    \item Variations in the geometry and cross-section properties of the K-braces along the length of the bridge are not taken into account.
    \item Shear lag in the deck is not modelled.
\end{itemize}

The uncertainty in the physical model prediction resulting from the aforementioned simplifications and misspecifications is accounted for in the Bayesian inference by considering the additive and multiplicative error models, introduced in Section \ref{section:probabilistic_models}.

\subsection{Correlation functions}
\label{section:kernels}

Correlation functions, also referred to as kernels or kernel functions in the literature and throughout this work, are positive definite functions of two Euclidean vectors $k(\bm{x},\bm{x}';\bm{\theta}_c)$ \citep{Duvenaud2014} that describe the correlation between points $\bm{x}$ and $\bm{x}'$. \autoref{tab:covariance_functions} provides a summary of the kernel functions used to model the correlation in the model prediction error in this paper. These kernel functions were chosen due to their wide adoption in statistical applications, ease of implementation and small number of parameters. Additionally, these kernel functions were empirically found to result in more accurate posterior and posterior predictive distributions among a group of candidate kernel functions in \cite{Koune2021}.

\begin{table}[htb!]
\centering
\caption{List of correlation functions and corresponding parameters.}
\label{tab:covariance_functions}
{\renewcommand{\arraystretch}{1.7}
\begin{tabular}{@{}lllll@{}}
\toprule
Kernel & Shorthand   & $k(\bm{x}, \bm{x}')$ & \multicolumn{2}{l}{Parameters} \\ 
\midrule
Independent & IID          & $\begin{cases} 1 & \text{if $\bm{x} = \bm{x'}$} \\ 0 & \text{if $\bm{x} \neq \bm{x'}$} \end{cases}$ & - \\
Radial Basis & RBF         & $\exp{(-\frac{\| \bm{x}, \bm{x'} \|^2}{2 l_{\mathrm{corr}}^2})}$   & $l_{\mathrm{corr}}$ \\
Exponential & EXP         & $\exp{\left( \frac{\| \bm{x}, \bm{x'} \|}{l_{\mathrm{corr}}} \right)}$ & $l_{\mathrm{corr}}$ \\
\bottomrule
\end{tabular}
}
\end{table}

\subsection{Physical model parameters}

The set of physical model parameters to be inferred, $\bm{\theta}_\mathrm{s}$, is shown in \autoref{fig:model_and_LLF}. The choice of the uncertain physical model parameters is based primarily on sensitivity analyses, the damage mechanisms expected to affect the behaviour of the structure and consultation with steel bridge experts. Independent rotational springs are defined at supports F, G, H and J, with the corresponding rotational spring stiffnessses denoted as $K_{\mathrm{r,1}}$ through $K_{\mathrm{r,4}}$, to simulate the friction at the support bearings and to account for the possibility of partial locking. An analysis of the sensitivity of the stress response to the stiffness of the support bearings indicated that the stiffness at supports K and L has negligible influence on the stress at the sensor locations. Therefore, the bearings at supports K and L are assumed to be hinges with no rotational stiffness. Furthermore, the riveted connections present in the K-braces result in high uncertainty on their stiffness, while additionally this stiffness was found to have a significant effect on the FE model response. The stiffness parameter of the vertical springs, $K_{\mathrm{v}}$, is assumed to be equal for all vertical springs along the length of the bridge. Both the vertical and rotational spring stiffness values span several orders of magnitude. We therefore represent these parameters in terms of their base-10 logarithms. In addition to the more natural and convenient representation, the parametrization by the base-10 logarithm yields a reduction of the relative size of the prior to the posterior, and therefore a reduction of the number of samples required for convergence when using nested sampling.

\autoref{tab:priors_structural} summarizes the prior distributions that are used for $\mathrm{log}_{10}(K_{\mathrm{r}})$ and $\mathrm{log}_{10}(K_{\mathrm{v}})$. The prior distributions are determined using a combination of engineering judgement and sensitivity analysis. Additional details of the sensitivity analysis for the rotational and vertical spring stiffness parameters can be found in Appendix \ref{appendix:sensitivity}. It should be noted that the impact of the physical model parameterization on the Bayesian model selection is not considered in this study. The feasibility of inferring parameters of the probabilistic model, and selecting the most likely model from a pool of candidate models, when a large number of physical model parameters is present is left as a potential topic for future work.

\begin{table}[htb!]
\centering
\caption{Description and uniform prior distribution bounds of physical model parameters.}
\label{tab:priors_structural}
\begin{tabular}{@{}llLll@{}}
\toprule
 Parameter & Unit & Description & Lower bound & Upper bound                          \\ \midrule
$\mathrm{log_{10}}(K_\mathrm{r,1-4})$ & kNm/rad & Rotational spring stiffness at the supports F to J (see \autoref{fig:model_and_LLF}). & 4.0 & 10.0  \\
$\mathrm{log_{10}}(K_\mathrm{v})$ & kN/m & Stiffness of the vertical translational springs, representing the K-braces. & 0.0 & 8.0   \\ \bottomrule
\end{tabular}
\end{table}

\subsection{Probabilistic model parameters}
\label{section:probabilistic_model_parameters}
\autoref{tab:priors_uncertainty} provides an overview of the set of probabilistic model parameters to be inferred, $\bm{\theta}_\mathrm{c}$, and their prior distributions. Details of the probabilistic model formulations for each of the cases investigated are provided in Sections \ref{section:case_1} and \ref{section:case_2}. The support (domain) of the priors must be defined such that it is possible to capture the structure of the correlations in the residuals between model predictions and measurements. It should be noted that choosing the priors for the parameters of the probabilistic model is not a simple task as no information on the correlation structure is available. Furthermore, a poor choice of prior can significantly impact the inference and prediction, leading to wide credible intervals \citep{Fuglstad2018}. Uniform priors are chosen with supports that are wide enough to capture a range of correlations that are expected to be present in the measurements.

\begin{table}[htb!]
\centering
\caption{Description and uniform prior distribution bounds of probabilistic model parameters.}
\label{tab:priors_uncertainty}
\begin{tabular}{@{}llLll@{}}
\toprule
Parameter & Unit & Description                                                                                                             & Lower bound & Upper bound \\ \midrule
$C_{\mathrm{v}}$ & [-] & COV of the multiplicative model prediction error. & 0.0         & 1.0         \\
$\sigma_{\mathrm{model}}$ & MPa   & Standard deviation of the additive model prediction error.               & 0.0         & 5.0         \\
$\sigma_{\mathrm{meas}}$ & MPa   & Standard deviation of the additive measurement error.                         & 0.0         & 1.0         \\
$l_{\mathrm{corr,t}}$ & m   & Temporal correlation lengthscale.                                                                                        & 0.0         & 300.0       \\
$l_{\mathrm{corr,x}}$ & m   & Spatial correlation lengthscale.                                                                                         & 0.0         & 300.0       \\ \bottomrule
\end{tabular}
\end{table}

\subsection{Notation}

The coupled probabilistic-physical models considered in the case studies presented in Sections \ref{section:case_1} and \ref{section:case_2} are distinguished by the model prediction error which can be multiplicative (\autoref{eq:multiplicative_model}) or additive (\autoref{eq:additive_model}), the kernel function used in the probabilistic model and the number of measurements used in the inference per influence line. A shorthand notation will be adopted to refer to the models for the remainder of this paper. Models will be referred to by the kernel that describes the temporal correlation (as shown in \autoref{tab:covariance_functions}) and the type of model prediction error considered, i.e. multiplicative or additive, denoted by the suffixes "M" and "A" respectively. As an example, following this notation, a model with multiplicative error where complete independence between the errors is considered will be written as IID-M.



\section{Exploratory analyses on the IJsselbridge case study using synthetic measurements}
\label{section:case_1}
\subsection{Description}

Initially, a synthetic example is studied to explore the impact of the size of the dataset on the feasibility of inferring the functional form the probabilistic model and the posterior distribution of the uncertain parameters. A pool of candidate probabilistic models with different correlation functions is formed, as described in \autoref{tab:summary_models_case_1}. For each probabilistic model a number of synthetic datasets with varying size (i.e. varying spatial and temporal discretization) are generated by evaluating the response of the physical model, and contaminating this response with random samples drawn from the probabilistic model for a prescribed set of ground truth values of the parameters. Bayesian inference is then performed for all models for each dataset, using the nested sampling method described in Section \ref{section:continuous_bayes} to estimate the posterior distribution and evidence. The resulting point estimates of the parameters of the probabilistic model are compared to the ground truth, under the assumption that the probabilistic model used to generate each dataset is known a priori. Additionally, the most likely model corresponding to each dataset is determined by comparing the estimated evidence under each probabilistic model for different dataset sizes. The aim of this synthetic case study is twofold:

\begin{itemize}
    \item To investigate the impact of the size of the dataset, and the functional form of the probabilistic model, on the accuracy of maximum a posteriori (MAP) estimates of the parameters of the probabilstic model.
    \item To gain insight into the size of the dataset required to correctly identify the probabilistic model from a pool of candidate models.
\end{itemize}

\begin{figure}[htb!]
     \centering
     \begin{subfigure}[b]{0.32\textwidth}
         \centering
         \includegraphics[width=\textwidth]{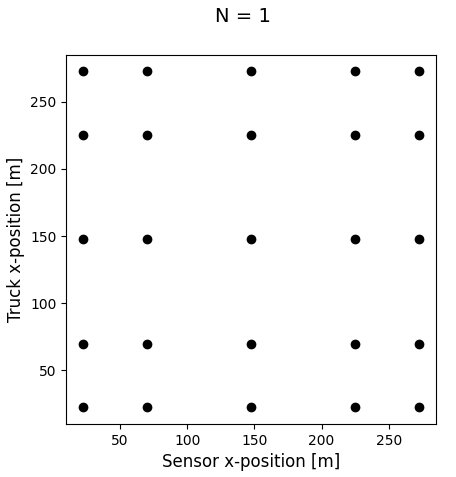}
         \label{fig:synthetic_example_grid_shape_1}
     \end{subfigure}
     \hfill
     \begin{subfigure}[b]{0.32\textwidth}
         \centering
         \includegraphics[width=\textwidth]{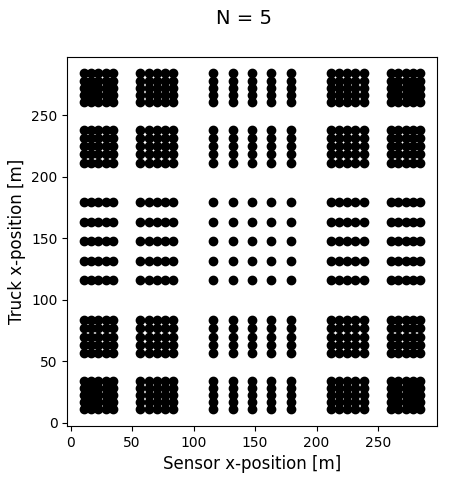}
         \label{fig:synthetic_example_grid_shape_2}
     \end{subfigure}
          \hfill
     \begin{subfigure}[b]{0.32\textwidth}
         \centering
         \includegraphics[width=\textwidth]{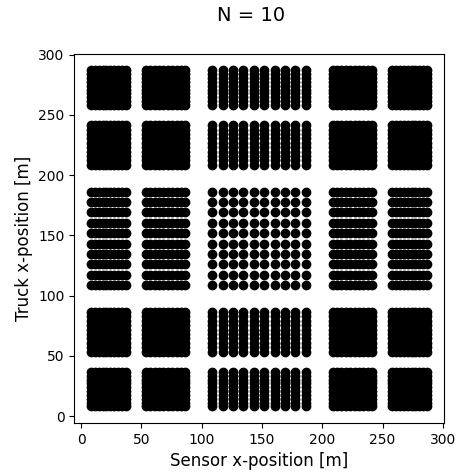}
         \label{fig:synthetic_example_grid_shape_3}
     \end{subfigure}
     
     \caption{Rectilinear grid of sensor and measurement positions considered in the synthetic case study for $N_x = N_t = \{ 1, 5, 10 \}$.}
     \label{fig:synthetic_example_grid_shape}
 \end{figure}

The example is structured as follows: The response of the physical model is evaluated for a set of ground truth parameters for increasing numbers of sensors per span, and considering an equal number of measurements in time. Denoting the number of sensors per span with $N_{\mathrm{x}}$ and the number of points per influence line by $N_{\mathrm{t}}$ (with each sensor yielding one influence line for each of the two traffic lanes), the resulting physical model response is a rectilinear grid with $N_{\mathrm{x}} = N_{\mathrm{t}}$. In this manner, the physical model response is evaluated for $N_{\mathrm{x}} = \{1, 2, ... 10\}$ sensors per span, with each sensor yielding two influence lines, each with $N_{\mathrm{t}} = \{1, 2, ... 10\}$ measurements respectively. The sensors are placed such that they are equally spaced within each span, with the distance between sensors taken as $L_{\mathrm{span}}/(N_{\mathrm{x}}+1)$ (\autoref{fig:synthetic_example_grid_shape}), where $L_{\mathrm{span}}$ denotes the length of the span. The physical model response is then contaminated with noise samples drawn from the probabilistic models summarized in \autoref{tab:summary_models_case_1}. To reduce the impact of the random sampling of the noise on the synthetic case study results, 50 samples are generated from each probabilistic model and for each grid size, and the resulting MAP estimates and the evidence for each model are averaged over the samples. The number of simulations was chosen as a compromise between having a large enough sample size to minimize the effects of the random sampling in the synthetic dataset, and the computational cost of performing multiple Bayesian inference realizations for a large number of models over a range of grid sizes.

\begin{table}[htb!]
\centering
\caption{Overview of models used in the case with synthetic measurements. See \autoref{tab:covariance_functions} for the meaning of the abbreviations and the details of the correlation function.}
\label{tab:summary_models_case_1}
\begin{tabular}{@{}lllll@{}}
\toprule
Model    & Shorthand & \begin{tabular}[c]{@{}c@{}}Temporal\\ correlation\end{tabular}   & \begin{tabular}[c]{@{}c@{}}Spatial\\ correlation\end{tabular} & $\bm{\theta}_\mathrm{c}$         \\ \midrule
$\mathcal{M}_1$    &    IID-M   & Independent           & Independent                   & $C_{\mathrm{v}}$, $\sigma_{\mathrm{meas}}$               \\
$\mathcal{M}_2$     &    RBF-M  & Radial Basis             & Exponential                  & $C_{\mathrm{v}}$, $\sigma_{\mathrm{meas}}$, $l_{\mathrm{corr,x}}$, $l_{\mathrm{corr,t}}$        \\
$\mathcal{M}_3$     &    EXP-M  & Exponential             & Exponential               & $C_{\mathrm{v}}$, $\sigma_{\mathrm{meas}}$, $l_{\mathrm{corr,x}}$, $l_{\mathrm{corr,t}}$          \\
$\mathcal{M}_4$    &    IID-A   & Independent         & Independent                  & $\sigma_{\mathrm{model}}$               \\
$\mathcal{M}_5$     &    RBF-A  & Radial basis             & Exponential                   & $\sigma_{\mathrm{model}}$, $\sigma_{\mathrm{meas}}$, $l_{\mathrm{corr,x}}$, $l_{\mathrm{corr,t}}$        \\
$\mathcal{M}_6$     &    EXP-A  &  Exponential &          Exponential                & $\sigma_{\mathrm{model}}$, $\sigma_{\mathrm{meas}}$, $l_{\mathrm{corr,x}}$, $l_{\mathrm{corr,t}}$          \\
\bottomrule
\end{tabular}
\end{table}

\subsection{Results}

\begin{figure}[htb!]
     \centering
     
     \begin{subfigure}[b]{0.49\textwidth}
         \centering
         \FIG{\includegraphics[width=0.9\textwidth]{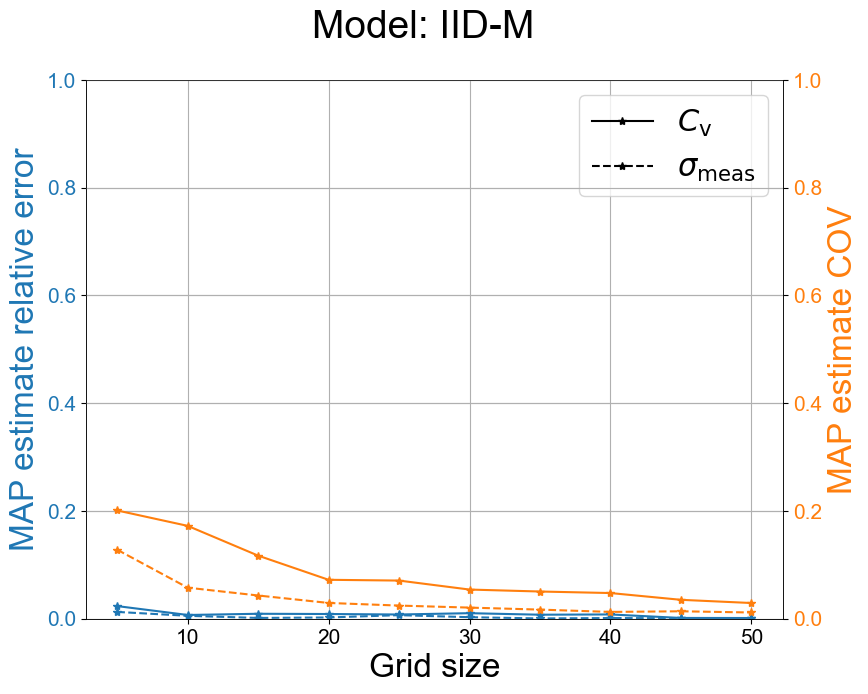}}
         \label{fig:synthetic_IID_m}
     \end{subfigure}
     \hfill
     \begin{subfigure}[b]{0.49\textwidth}
         \centering
         \FIG{\includegraphics[width=0.9\textwidth]{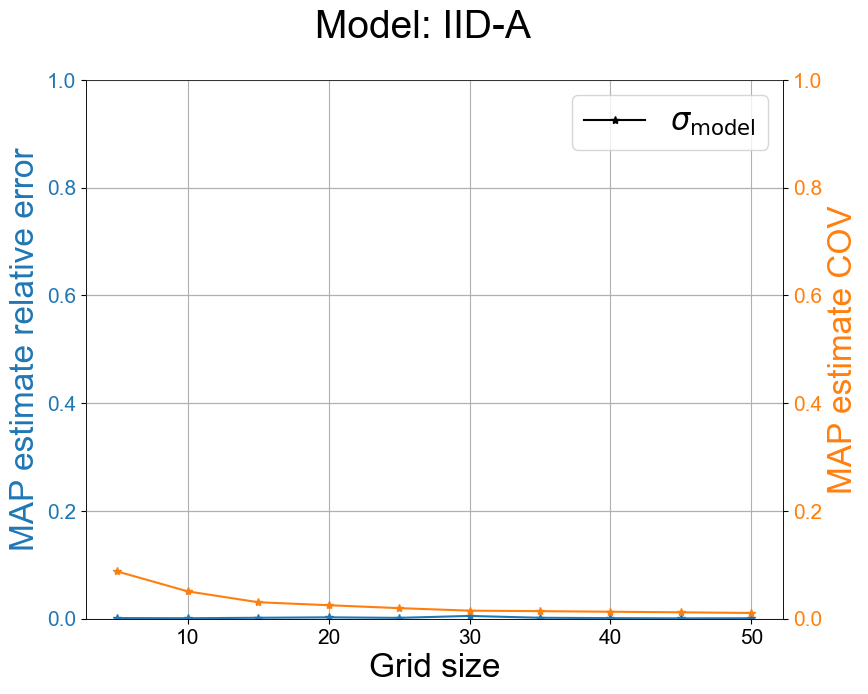}}
         \label{fig:synthetic_IID_a}
     \end{subfigure}
     
     \hfill
     
     \begin{subfigure}[b]{0.49\textwidth}
         \centering
         \FIG{\includegraphics[width=0.9\textwidth]{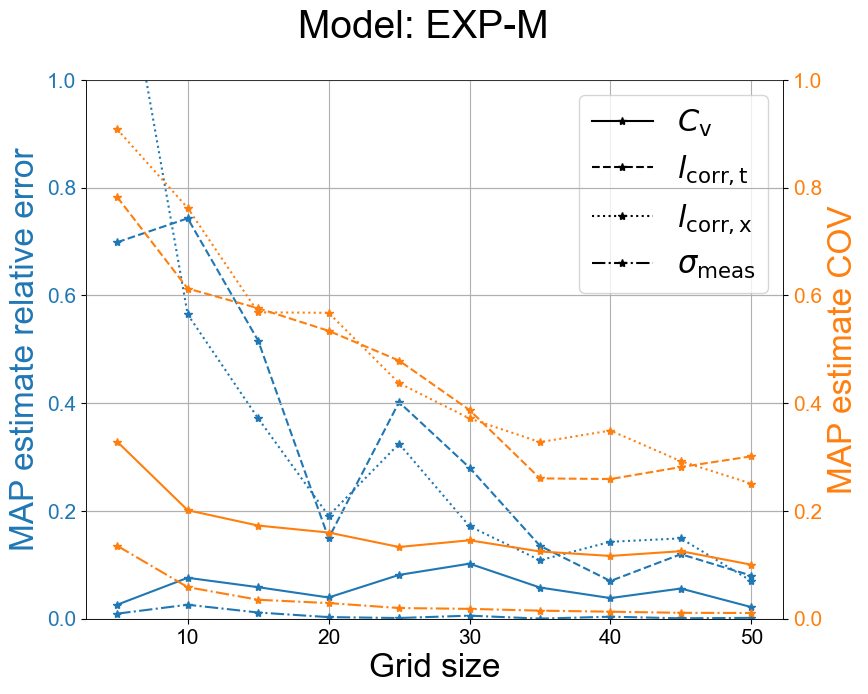}}
         \label{fig:synthetic_EXP_m}
     \end{subfigure}
     \hfill
     \begin{subfigure}[b]{0.49\textwidth}
         \centering
         \FIG{\includegraphics[width=0.9\textwidth]{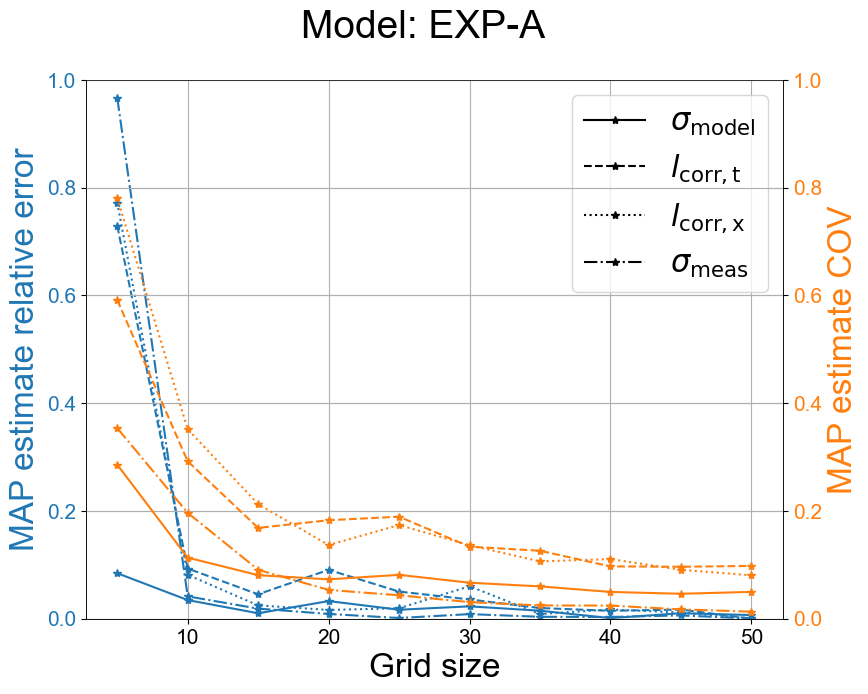}}
         \label{fig:synthetic_EXP_a}
     \end{subfigure}
     
     \hfill
     
     \begin{subfigure}[b]{0.49\textwidth}
         \centering
         \FIG{\includegraphics[width=0.9\textwidth]{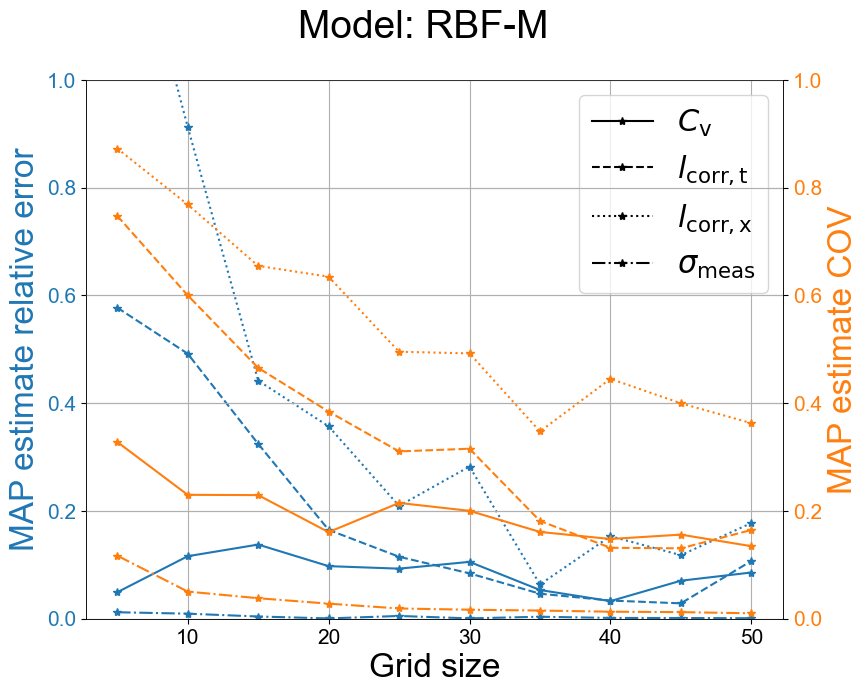}}
         \label{fig:synthetic_RBF_m}
     \end{subfigure}
     \hfill
     \begin{subfigure}[b]{0.49\textwidth}
         \centering
         \FIG{\includegraphics[width=0.9\textwidth]{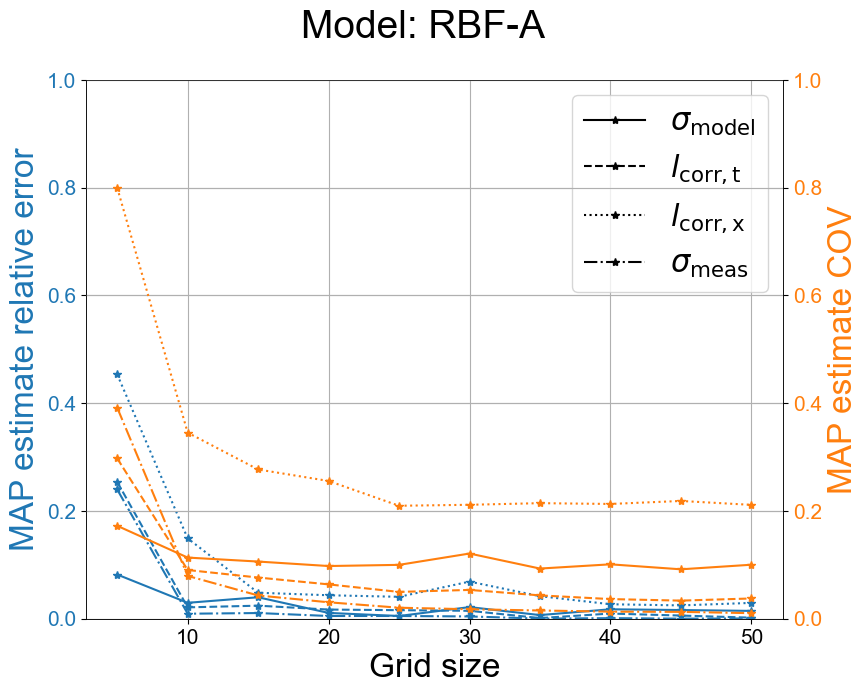}}
         \label{fig:synthetic_RBF_a}
     \end{subfigure}
     
        \caption{Relative error of the mean MAP estimates of probabilistic model parameters compared to the ground truth, and COV of MAP estimates as a function of grid size.}
        \label{fig:synthetic_case_MAP_estimates}
\end{figure}

Bayesian inference is performed for the models $\mathcal{M}_1$ to $\mathcal{M}_6$ listed in \autoref{tab:summary_models_case_1}. At each grid size, we examine if the correlation parameters are correctly inferred, when performing inference with the true probabilistic model used to generate the measurements, by computing the relative error between the known ground truth for each probabilistic model and the mean MAP point estimate obtained from the Bayesian inference. The scatter of the MAP estimates for each ground truth model, expressed in terms of the COV is also computed in order to quantify the impact of the random sampling as a function of the grid size. The results are shown in \autoref{fig:synthetic_case_MAP_estimates}. A grid size of $10 \times 10$ (corresponding to $N=2$ sensors per span) is sufficient to obtain an accurate point estimate of the multiplicative model prediction uncertainty COV parameter $C_{\mathrm{v}}$, the standard deviation of the additive model prediction uncertainty $\sigma_{\mathrm{model}}$ and the standard deviation of the measurement uncertainty $\sigma_{\mathrm{meas}}$ (i.e. the relative error is below 0.10). For models with multiplicative model prediction uncertainty, both the relative error and the COV of the MAP estimate of the spatial and temporal correlation length parameters are heavily dependent on the grid size. This is not the case for models with additive prediction uncertainty, where the relative error is not significantly affected when increasing the grid resolution beyond $10 \times 10$. This could potentially be explained by considering that the multiplicative error structure introduces a dependency of the probabilistic model on the physical model. We speculate that this additional complexity may hinder the estimation of the posterior distribution of the parameters of multiplicative models, resulting in slower convergence (with respect to the grid size) of the point estimates of the probabilistic model parameters to the ground truth values. It is noted that the accuracy of the obtained point estimates for the correlation length parameters is also expected to depend on their size relative to the distances between points on the measurement grid. However, this dependence is not further examined in this paper.

By calculating the log-evidence for each of the models $\mathcal{M}_1$ to $\mathcal{M}_6$ we investigate if the correct probabilistic model is identified from the pool of candidate models for each dataset. The log of the average evidence -- based on 50 generated datasets -- obtained for each probabilistic model and for each ground truth model and grid size is shown in \autoref{tab:synthetic-case-evidence}. The correct ground truth model is identified with a single sensor per span in the case of additive model prediction uncertainty. For multiplicative models, the ground truth is correctly identified for two to three sensors per span.

Additionally, the posterior probability of the ground truth model $p_{\mathrm{gt}}$, and the identification accuracy (i.e. the percentage of realizations for which the ground truth model obtains the highest posterior probability) for increasing refinement of the grid of sensors are also shown in \autoref{tab:synthetic-case-evidence}. The posterior probabilities are obtained using \autoref{eq:bayes_model_select}, with the evidence for each model taken as the average across the independent realizations. It is observed that multiplicative models generally require finer discretization and larger datasets, compared to additive models, to achieve similar levels of accuracy. For the additive models, the ground truth model is recovered with perfect accuracy for three or more sensors per span, with the exception of the IID-A case. The comparatively low posterior probability of the ground truth model observed for the IID-A case can be attributed to the fact that all of the considered models are able to describe uncorrelated additive Gaussian noise.

                                          
\begin{table}[!p]
\caption{Log of the mean evidence, posterior probability of the ground truth model, and identification accuracy per model as a function of the number of sensors per span for different ground truth models, averaged over 50 randomly generated datasets. Underscores denote the ground truth model used to generate the data and bold type denotes the highest evidence.}
\label{tab:synthetic-case-evidence}
\resizebox{\textwidth}{!}{%
\begin{tabular}{@{}lllllllllll@{}}
\toprule
N                                  & 1                & 2                & 3                 & 4                 & 5                 & 6                 & 7                 & 8                 & 9                 & 10                 \\ \midrule
\multicolumn{1}{l|}{{\ul IID-M}}   & -66.50           & \textbf{-267.45} & \textbf{-587.77}  & \textbf{-1050.10} & \textbf{-1635.36} & \textbf{-2357.95} & \textbf{-3198.63} & \textbf{-4210.59} & \textbf{-5329.55} & \textbf{-6539.73}  \\
\multicolumn{1}{l|}{RBF-M}         & \textbf{-66.08}  & -273.26          & -594.69           & -1057.69          & -1644.28          & -2365.88          & -3207.05          & -4220.28          & -5340.00          & -6550.03           \\
\multicolumn{1}{l|}{EXP-M}         & -66.44           & -272.45          & -594.31           & -1057.76          & -1644.18          & -2365.74          & -3206.78          & -4219.59          & -5340.46          & -6549.96           \\
\multicolumn{1}{l|}{IID-A}         & -74.11           & -292.36          & -634.86           & -1156.65          & -1758.60          & -2602.01          & -3562.04          & -4604.75          & -5903.94          & -7294.54           \\
\multicolumn{1}{l|}{RBF-A}         & -76.82           & -294.67          & -638.15           & -1160.86          & -1764.36          & -2609.82          & -3564.69          & -4608.00          & -5913.43          & -7302.32           \\
\multicolumn{1}{l|}{EXP-A}         & -76.87           & -294.77          & -637.77           & -1160.42          & -1764.56          & -2609.67          & -3564.56          & -4607.45          & -5913.02          & -7302.19           \\ \midrule
\multicolumn{1}{l|}{$p_\mathrm{gt}$}  & 0.28             & 0.99             & 1.00              & 1.00              & 1.00              & 1.00              & 1.00              & 1.00              & 1.00              & 1.00               \\
\multicolumn{1}{l|}{Acc.} & 0.62             & 0.96             & 1.00              & 1.00              & 1.00              & 1.00              & 1.00              & 1.00              & 1.00              & 1.00               \\ \midrule
\multicolumn{1}{l|}{IID-M}         & -72.97           & -262.56          & -569.84           & -1024.87          & -1624.73          & -2283.02          & -3133.34          & -4035.82          & -5174.47          & -6323.96           \\
\multicolumn{1}{l|}{{\ul RBF-M}}   & \textbf{-71.06}  & -260.07          & \textbf{-547.66}  & \textbf{-968.03}  & \textbf{-1536.89} & \textbf{-2159.71} & \textbf{-2947.94} & \textbf{-3854.52} & \textbf{-4875.21} & \textbf{-6022.68}  \\
\multicolumn{1}{l|}{EXP-M}         & -71.61           & \textbf{-259.56} & -549.72           & -970.29           & -1542.18          & -2162.22          & -2956.65          & -3864.16          & -4886.69          & -6027.02           \\
\multicolumn{1}{l|}{IID-A}         & -80.98           & -270.99          & -594.33           & -1091.26          & -1701.98          & -2344.76          & -3233.32          & -4073.89          & -5341.58          & -6585.27           \\
\multicolumn{1}{l|}{RBF-A}         & -82.65           & -275.41          & -598.35           & -1080.67          & -1673.37          & -2303.21          & -3182.04          & -4018.69          & -5187.90          & -6314.14           \\
\multicolumn{1}{l|}{EXP-A}         & -82.50           & -275.12          & -598.38           & -1082.46          & -1674.78          & -2307.78          & -3185.81          & -4021.77          & -5201.11          & -6327.58           \\ \midrule
\multicolumn{1}{l|}{$p_\mathrm{gt}$}  & 0.58             & 0.36             & 0.89              & 0.91              & 0.99              & 0.92              & 1.00              & 1.00              & 1.00              & 0.99               \\
\multicolumn{1}{l|}{Acc.} & 0.44             & 0.72             & 0.86              & 0.92              & 0.88              & 0.96              & 0.96              & 0.98              & 1.00              & 1.00               \\ \midrule
\multicolumn{1}{l|}{IID-M}         & -61.64           & -266.86          & -590.31           & -1033.72          & -1619.24          & -2317.31          & -3158.95          & -4162.92          & -5233.14          & -6482.46           \\
\multicolumn{1}{l|}{RBF-M}         & \textbf{-59.92}  & -263.45          & -579.58           & -999.59           & -1550.39          & -2227.75          & -3030.65          & -3944.78          & -5015.33          & -6114.11           \\
\multicolumn{1}{l|}{{\ul EXP-M}}   & -60.67           & \textbf{-262.29} & \textbf{-575.65}  & \textbf{-996.73}  & \textbf{-1541.55} & \textbf{-2216.66} & \textbf{-3018.81} & \textbf{-3918.58} & \textbf{-4975.22} & \textbf{-6090.35}  \\
\multicolumn{1}{l|}{IID-A}         & -67.82           & -296.88          & -614.16           & -1092.10          & -1689.28          & -2426.96          & -3270.71          & -4385.12          & -5494.79          & -6846.72           \\
\multicolumn{1}{l|}{RBF-A}         & -70.18           & -292.18          & -613.42           & -1089.07          & -1684.85          & -2377.04          & -3216.52          & -4245.98          & -5317.11          & -6551.88           \\
\multicolumn{1}{l|}{EXP-A}         & -70.26           & -291.23          & -613.26           & -1088.96          & -1684.45          & -2378.49          & -3214.03          & -4250.08          & -5321.82          & -6551.58           \\ \midrule
\multicolumn{1}{l|}{$p_\mathrm{gt}$}  & 0.29             & 0.76             & 0.98              & 0.95              & 1.00              & 1.00              & 1.00              & 1.00              & 1.00              & 1.00               \\
\multicolumn{1}{l|}{Acc.} & 0.30             & 0.82             & 0.98              & 0.90              & 0.98              & 0.98              & 1.00              & 1.00              & 1.00              & 1.00               \\ \midrule
\multicolumn{1}{l|}{IID-M}         & -110.91          & -457.77          & -1032.15          & -1822.18          & -2882.95          & -4150.56          & -5678.20          & -7375.49          & -9373.51          & -11566.59          \\
\multicolumn{1}{l|}{RBF-M}         & -110.69          & -457.87          & -1033.20          & -1822.45          & -2885.07          & -4150.93          & -5678.80          & -7376.49          & -9373.58          & -11566.89          \\
\multicolumn{1}{l|}{EXP-M}         & -110.84          & -457.80          & -1033.07          & -1822.57          & -2885.07          & -4150.86          & -5678.61          & -7376.35          & -9373.13          & -11566.89          \\
\multicolumn{1}{l|}{{\ul IID-A}}   & \textbf{-108.57} & \textbf{-454.82} & \textbf{-1029.84} & \textbf{-1819.42} & \textbf{-2881.06} & \textbf{-4147.81} & \textbf{-5674.38} & \textbf{-7371.99} & \textbf{-9369.32} & \textbf{-11563.00} \\
\multicolumn{1}{l|}{RBF-A}         & -110.11          & -457.30          & -1031.99          & -1822.60          & -2885.24          & -4151.45          & -5678.68          & -7376.64          & -9371.79          & -11566.81          \\
\multicolumn{1}{l|}{EXP-A}         & -109.91          & -457.29          & -1031.70          & -1822.44          & -2885.15          & -4151.30          & -5678.59          & -7376.45          & -9371.84          & -11566.58          \\ \midrule
\multicolumn{1}{l|}{$p_\mathrm{gt}$}  & 0.56             & 0.76             & 0.69              & 0.80              & 0.82              & 0.82              & 0.93              & 0.93              & 0.82              & 0.89               \\
\multicolumn{1}{l|}{Acc.} & 0.80             & 0.92             & 0.92              & 0.94              & 0.96              & 0.94              & 1.00              & 0.96              & 1.00              & 0.98               \\ \midrule
\multicolumn{1}{l|}{IID-M}         & -116.12          & -449.47          & -1004.21          & -1773.42          & -2680.87          & -3635.65          & -5189.07          & -6853.51          & -8621.59          & -10818.35          \\
\multicolumn{1}{l|}{RBF-M}         & -116.25          & -449.20          & -1003.54          & -1727.56          & -2590.33          & -3540.85          & -5144.94          & -6612.07          & -8247.39          & -9938.49           \\
\multicolumn{1}{l|}{EXP-M}         & -116.22          & -449.45          & -1003.71          & -1743.65          & -2621.39          & -3569.22          & -5165.94          & -6711.81          & -8354.34          & -10122.47          \\
\multicolumn{1}{l|}{IID-A}         & -113.77          & -446.42          & -1000.72          & -1770.73          & -2681.02          & -3635.19          & -5184.55          & -6849.24          & -8619.79          & -10818.07          \\
\multicolumn{1}{l|}{{\ul RBF-A}}   & \textbf{-102.52} & \textbf{-327.30} & \textbf{-693.87}  & \textbf{-1171.40} & \textbf{-1758.91} & \textbf{-2431.99} & \textbf{-3295.85} & \textbf{-4260.14} & \textbf{-5318.40} & \textbf{-6524.88}  \\
\multicolumn{1}{l|}{EXP-A}         & -104.22          & -336.12          & -709.04           & -1197.58          & -1795.79          & -2490.29          & -3375.42          & -4337.38          & -5441.78          & -6677.48           \\ \midrule
\multicolumn{1}{l|}{$p_\mathrm{gt}$}  & 0.85             & 1.00             & 1.00              & 1.00              & 1.00              & 1.00              & 1.00              & 1.00              & 1.00              & 1.00               \\
\multicolumn{1}{l|}{Acc.} & 0.76             & 0.98             & 1.00              & 1.00              & 1.00              & 1.00              & 1.00              & 1.00              & 1.00              & 1.00               \\ \midrule
\multicolumn{1}{l|}{IID-M}         & -114.34          & -442.17          & -996.64           & -1777.73          & -2726.68          & -3959.78          & -5384.25          & -7211.06          & -9076.12          & -11267.86          \\
\multicolumn{1}{l|}{RBF-M}         & -114.40          & -442.10          & -996.48           & -1775.52          & -2707.97          & -3911.51          & -5211.81          & -7000.00          & -8721.13          & -10778.94          \\
\multicolumn{1}{l|}{EXP-M}         & -114.48          & -441.95          & -996.60           & -1775.57          & -2715.95          & -3918.94          & -5241.51          & -7056.89          & -8805.41          & -10875.05          \\
\multicolumn{1}{l|}{IID-A}         & -111.73          & -439.29          & -993.23           & -1774.09          & -2723.19          & -3956.71          & -5381.91          & -7207.27          & -9076.46          & -11280.75          \\
\multicolumn{1}{l|}{RBF-A}         & -110.85          & -393.23          & -837.71           & -1432.90          & -2120.02          & -3013.09          & -4039.86          & -5175.60          & -6460.53          & -7814.57           \\
\multicolumn{1}{l|}{{\ul EXP-A}}   & \textbf{-110.56} & \textbf{-389.56} & \textbf{-814.55}  & \textbf{-1383.96} & \textbf{-2068.01} & \textbf{-2906.41} & \textbf{-3836.56} & \textbf{-4944.19} & \textbf{-6174.49} & \textbf{-7484.61}  \\ \midrule
\multicolumn{1}{l|}{$p_\mathrm{gt}$}  & 0.47             & 0.98             & 1.00              & 1.00              & 1.00              & 1.00              & 1.00              & 1.00              & 1.00              & 1.00               \\
\multicolumn{1}{l|}{Acc.}                     & 0.54             & 0.92             & 1.00              & 1.00              & 1.00              & 1.00              & 1.00              & 1.00              & 1.00              & 1.00               \\ \bottomrule
\end{tabular}%
}
\end{table}

\subsection{Conclusions}
Based on the results presented previously it can be seen that, although the estimated evidence for each model is sensitive to the size of the dataset, the correct probabilistic model used to generate the data can be identified in all cases, for as few as three sensors per span. Furthermore, it is evident that the correlation structure and the size of the dataset impact the accuracy of the posterior distributions and point estimates of the parameters describing the uncertainty. For additive probabilistic models with correlation, the MAP estimates of the parameters converge faster (i.e. for a smaller number of measurements) to the known ground truth values, compared to the multiplicative probabilistic models. 

\section{Analyses on the IJsselbridge case study using real-world measurements}
\label{section:case_2}

\subsection{Analysis considering a single sensor}
\label{section:case_2a}

 An initial analysis is performed considering data from the H4 sensor (see \autoref{fig:fugro_influence_lines}), with the aim of assessing the feasibility of performing system identification while considering dependencies in the model prediction uncertainties, and to examine the benefit of the additional data compared to a reference case where only four hand-picked measurements from the largest peaks and troughs of each influence line are considered. A second analysis with multiple sensors is performed to demonstrate the feasibility of considering large datasets under combined spatial and temporal dependencies and to determine the efficiency of the block Cholesky log-likelihood evaluation presented in Section \ref{section:efficient_loglikelihood_multiplicative}.

\autoref{tab:summary_models_case_1} provides an overview of the considered models in the real-world case study, including their labels, correlation structure, dataset size and parameters. Compared to the synthetic case study, two ``reference" analyses are added, referred to as ``REF", in which we adopt a small dataset with $4 \times 2$ points. These analyses represent a typical application of Bayesian system identification for structures, where only a limited number of manually selected measurements are included in the dataset. In these reference analyses we assume that the discrepancies between measurement and model prediction are fully independent. The four largest (in absolute value) peaks per influence line are chosen to form this dataset, such that they maximize the amount of information regarding the parameters of interest while being spaced far enough apart to be considered independent. The assumption of independence is based mainly on engineering judgement, which is often the case in applications where a limited amount of measurements makes it infeasible to assess their independence.

\begin{table}[htb!]
\centering
\caption{Overview of models used in the case with real-world measurements. See \autoref{tab:covariance_functions} for the meaning of the abbreviations and the details of the correlation function.}
\label{tab:summary_models_case_2}
\begin{tabular}{@{}lllll@{}}
\toprule
Model    & Shorthand & Temporal correlation        & Dataset size\tablefootnote{The factor $2$ in the dataset size is included to indicate that each sensor yields two influence lines, one for each controlled loading test as discussed in Section \ref{section:fugro_measurements}.}        & $\bm{\theta}_\mathrm{c}$         \\ \midrule
$\mathcal{M}_1$     &    IID-M & Independent        & $193 \times 2$                    & $C_{\mathrm{v}}$, $\sigma_{\mathrm{meas}}$               \\
$\mathcal{M}_2$     &    RBF-M & Radial Basis       & $193 \times 2$                  & $C_{\mathrm{v}}$, $\sigma_{\mathrm{meas}}$, $l_{\mathrm{corr,t}}$        \\
$\mathcal{M}_3$     &    EXP-M & Exponential           & $193 \times 2$                & $C_{\mathrm{v}}$, $\sigma_{\mathrm{meas}}$, $l_{\mathrm{corr,t}}$          \\
$\mathcal{M}_4$     &    REF-M & Independent       & $4 \times 2$                    & $C_{\mathrm{v}}$, $\sigma_{\mathrm{meas}}$           \\
$\mathcal{M}_5$     &    IID-A & Independent        & $193 \times 2$                   & $\sigma_{\mathrm{model}}$               \\
$\mathcal{M}_6$     &    RBF-A & Radial Basis       & $193 \times 2$                   & $\sigma_{\mathrm{model}}$, $\sigma_{\mathrm{meas}}$, $l_{\mathrm{corr,t}}$        \\
$\mathcal{M}_7$     &    EXP-A & Exponential           & $193 \times 2$                & $\sigma_{\mathrm{model}}$, $\sigma_{\mathrm{meas}}$, $l_{\mathrm{corr,t}}$          \\
$\mathcal{M}_8$     &    REF-A & Independent        & $4 \times 2$                     & $\sigma_{\mathrm{model}}$               \\ 
\bottomrule
\end{tabular}
\end{table}

Inference is performed using the nested sampling technique, yielding an estimate of the posterior distribution and evidence for each of the models $\mathcal{M}_{1} - \mathcal{M}_{8}$. The IID-M, REF-M, IID-A and REF-A models consider complete independence in the model prediction uncertainty, while models RBF-M, EXP-M, RBF-A and EXP-A assume dependencies modeled by exponential and radial basis kernel functions. The uniform prior distributions assumed for the physical model parameters and uncertainty parameters are listed in \autoref{tab:priors_structural} and \autoref{tab:priors_uncertainty} respectively. It is noted that in models IID-M and IID-A, complete independence is assumed for the model prediction uncertainty, despite the dense spacing of the measurements. It is therefore expected that the uncertainty in the posterior distributions will be significantly underestimated if dependencies are present in the model prediction error. The posterior distributions obtained by this model are included for the purpose of comparing the inferred means of the parameters with other models that assume dependence, as well as illustrating the effect of increasing the number of points under the independence assumption on the posterior distributions.

\begin{figure}[htb!]
    \centering
    \FIG{\includegraphics[width=0.8\textwidth]{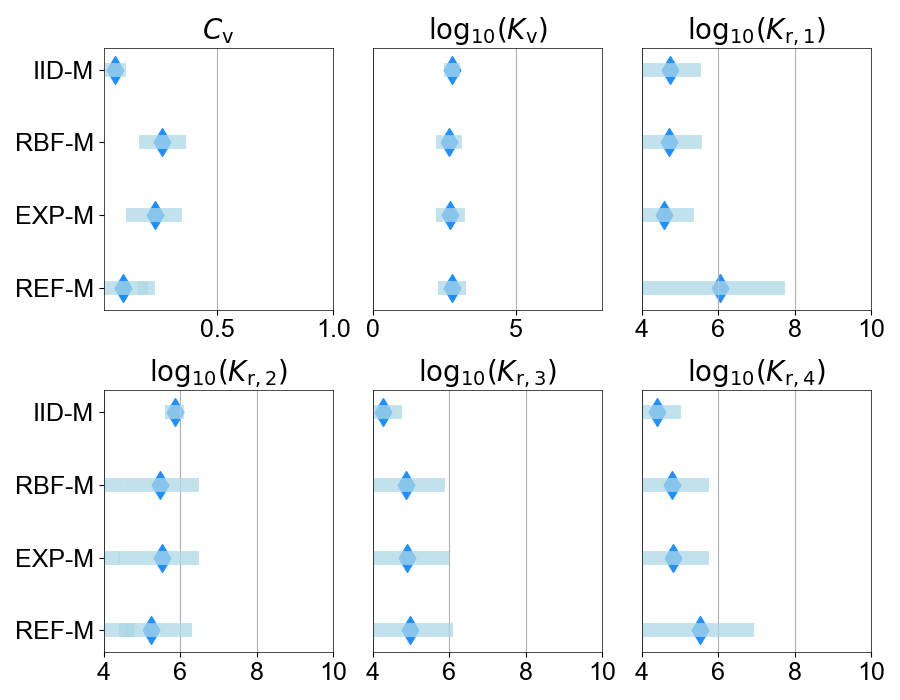}}
    \caption{Comparison of posterior mean and $90\%$ HD CIs for models with multiplicative uncertainty structure.}
    \label{fig:case_1_posterior_CIs_multiplicative}
\end{figure}

\begin{figure}[htb!]
    \centering
    \FIG{\includegraphics[width=0.8\textwidth]{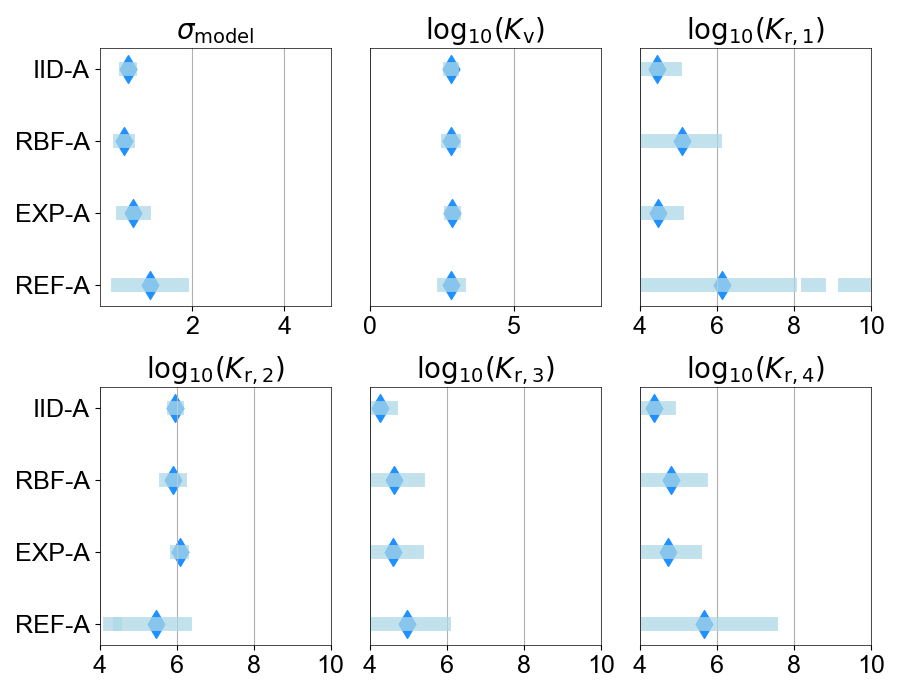}}
    \caption{Comparison of posterior mean and $90\%$ HD CIs for models with additive uncertainty structure.}
    \label{fig:case_1_posterior_CIs_additive}
\end{figure}

\autoref{fig:case_1_posterior_CIs_multiplicative} and \autoref{fig:case_1_posterior_CIs_additive} show the posterior distribution highest density (HD) credible intervals (CIs) for models with multiplicative model prediction uncertainty and additive model prediction uncertainty. Comparing the reference models $\mathcal{M}_{4}$ and $\mathcal{M}_{8}$ with the other models, we observe that for the additive probabilistic model the posterior CIs of the physical model parameters are wider, indicating that the additional information contained in the full dataset of $193 \times 2$ data points can result in reduced uncertainty in the posterior distributions of the parameters of interest. This can also be observed for parameters $\mathrm{log}_{10}(K_{\mathrm{r,1}})$ and $\mathrm{log}_{10}(K_{\mathrm{r,4}})$ in the case of multiplicative model prediction uncertainty. Furthermore, for the RBF-M and EXP-M models higher point estimates are obtained for $C_{\mathrm{v}}$ compared to both the IID-M and REF-M models, indicating that including correlation parameters in the vector of uncertain parameters to be inferred can affect the inference of other parameters of interest, or that it requires an increase in the size of the dataset to obtain accurate estimates and low uncertainty in the posterior distribution of parameters of the probabilistic model. 


The number of parameters considered in the Bayesian inference, and the probabilistic model of the data generating process used to obtain the likelihood function can have a significant impact on the number of likelihood function evaluations required to achieve convergence when performing Bayesian inference using nested sampling. The total number of function evaluations (NFE) per model is shown in the second column of \autoref{tab:autogenerated_model_selection_case_1}. A range of approximately $29\cdot 10^3$ to $109 \cdot 10^3$ likelihood evaluations are needed to achieve convergence depending on the model. The models where correlation is considered generally require a larger number of likelihood evaluations compared to models where the model prediction errors are taken as independent. Although no clear conclusions arise in this case regarding the impact of the number of parameters, correlation structure and dataset size on the NFE, it is found that the reference cases, and the cases where independence is assumed, generally require considerably fewer likelihood function evaluations compared to models where correlation is considered.

Nested sampling also yields a noisy estimate of the evidence, which can be used to determine the most likely model $\mathcal{M}_{\mathrm{i}}$ within the candidate pool of models. Based on these estimated evidences, the posterior probability of each model in $\bm{\mathcal{M}}$ is calculated using \autoref{eq:bayes_model_select}. All models in $\bm{\mathcal{M}}$ are considered a priori equally likely. \autoref{tab:autogenerated_model_selection_case_1} provides the log-evidence, posterior probability and interpretation of the Bayes factor per model. It is noted that the reference models $\mathcal{M}_{\mathrm{4}}$ and $\mathcal{M}_{\mathrm{8}}$ are not included in the model selection due to the different dataset used with these models, and therefore no evidence, posterior probabilities or Bayes factors are obtained. It is possible to observe that the assumption of complete independence in the model prediction uncertainty is not supported by the evidence, which is reflected in the low values of the log-evidence for the IID-M and IID-A models. The additive model with exponentially correlated model prediction uncertainty (EXP-A) is the most likely model, with practically $100\%$ posterior probability. This indicates that the data decisively supports the inclusion of correlation. It should be emphasized that the posterior probability only reflects the likelihood of a model compared to other models in the candidate pool, and cannot be used to draw conclusions on the validity of a model in general.

\begin{table}[htb!] 
        \centering 
        \caption{NFE required for convergence rounded to the nearest thousandth, log-evidence, posterior probability and Bayes factors per model. All models included in the model selection are assumed a priori equally likely.} 
        \label{tab:autogenerated_model_selection_case_1} 
\begin{tabular}{llllll}
\hline
 Model & Shorthand & NFE ($\times 1000$) &   log($\mathcal{Z}$) &  p($\mathcal{M}$) & Evidence against          \\
\hline
$\mathcal{M}_{\mathrm{1}}$ & IID-M   & 64 &              -358.28 &               0.00 & Decisive                \\
$\mathcal{M}_{\mathrm{2}}$ & RBF-M   & 93 &               -58.06 &               0.00 & Decisive                \\
$\mathcal{M}_{\mathrm{3}}$ & EXP-M   & 76 &              -126.95 &               0.00 & Decisive                \\
$\mathcal{M}_{\mathrm{4}}$ & REF-M   & 51 &               -      &                     -     &   -                     \\
$\mathcal{M}_{\mathrm{5}}$ & IID-A   & 41 &              -381.15 &               0.00 & Decisive                \\
$\mathcal{M}_{\mathrm{6}}$ & RBF-A   & 109 &               322.22 &               0.00  & Decisive                \\
$\mathcal{M}_{\mathrm{7}}$ & EXP-A   & 92 &               349.55 &               1.00  & Barely worth mentioning \\
$\mathcal{M}_{\mathrm{8}}$ & REF-A   & 29 &               -      &                     -     &   -                     \\
\hline
\end{tabular}
\end{table}

\begin{figure}[H]
     \centering
     \begin{subfigure}[b]{0.47\textwidth}
         \centering
         \includegraphics[width=\textwidth]{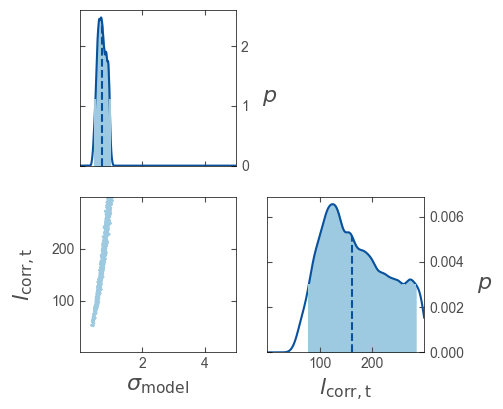}
         \label{fig:unidentifiability_exp_a}
     \end{subfigure}
     \hfill
     \begin{subfigure}[b]{0.49\textwidth}
         \centering
         \includegraphics[width=\textwidth]{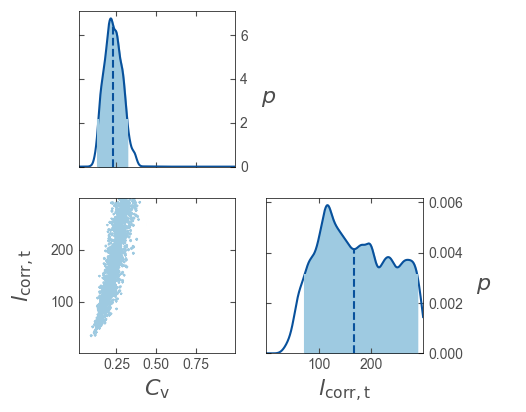}
         \label{fig:unidentifiability_exp_m}
     \end{subfigure}
    \caption{Joint unidentifiability of the correlation length and model prediction uncertainty parameters for the exponential kernel considering additive (left) and multiplicative (right) model prediction uncertainty.}
    \label{fig:unidentifiability}
\end{figure}

As mentioned in Section \ref{section:probabilistic_model_parameters}, the choice of prior distributions for the probabilistic model parameters can have a significant impact on the posterior and posterior predictive distributions. It was observed that the models with exponential correlation for the case of a single sensor are particularly sensitive to prior distribution of the correlation length parameter, due to the joint unidentifiability of the marginal variance and the correlation length parameters \citep{Fuglstad2018}. The corresponding joint posterior distributions for the EXP-M and EXP-A models are shown in \autoref{fig:unidentifiability}. It can be seen that specifying a large upper bound on the prior of the correlation length will result in wide credible intervals in the posteriors of the corresponding $\sigma_{\mathrm{model}}$ and $C_{\mathrm{v}}$ parameters. This effect is less pronounced in the case of multiple sensors presented in Section \ref{section:case_2b} and furthermore does not affect the RBF kernel in either the single or multiple sensor case.

\subsection{Analysis considering multiple sensors}
\label{section:case_2b}

In order to evaluate the feasibility of the approach under combined spatial and temporal dependencies, Bayesian inference is performed considering influence lines from multiple sensors. For the reference models REF-M and REF-A, four measurements are selected per influence line, with each sensor yielding two influence lines. Specifically, one peak stress measurement per span is selected from the four spans with the largest absolute peak stresses. For the remaining models, 193 measurements per influence line are considered from locations along each influence line corresponding to the longitudinal positions of nodes in the FE model. The dataset includes influence lines for truck trial runs on the left and right lane for sensors H1, H2, H4, H5, H9 and H10. The remaining sensor data exhibits structural behaviour that can not be captured by the twin girder FE model presented in Section \ref{section:physical_model}, and is therefore discarded.

\autoref{tab:summary_models_case_2} provides an overview of the considered models in the analyses using data from multiple sensors. In Section \ref{section:case_2a}, with data coming from one sensor we only dealt with temporal correlation. Here, both spatial and temporal correlation is present in the model prediction error. For all models in $\bm{\mathcal{M}}$ the temporal correlation is described by an exponential kernel function. The spatial correlation is only considered in the models $\mathcal{M}_{2}$, $\mathcal{M}_{3}$, $\mathcal{M}_{6}$, and $\mathcal{M}_{7}$, also using an exponential kernel function. In the other models, spatial correlation is not taken into account, with each strain gauge along the length of the bridge considered fully independent. By comparing the posterior model probabilities for models with and without spatial correlation, we can evaluate the validity of the spatial correlation assumption, and determine which description of the model prediction uncertainties is best supported by the data. 

The assumption of exponentially correlated model prediction uncertainty in space provides a computational advantage, compared to other kernel functions. Due to poor scaling of the computational complexity of the multivariate Gaussian log-likelihood with the size of the dataset, the use of a conventional approach to evaluate the log-likelihood (e.g. by factorizing the full covariance matrix) would result in significant computational cost even for datasets with a few thousand observations when correlation is taken into account. To alleviate this issue, the efficient log-likelihood evaluation approaches described in Section \ref{section:efficient_likelihood} are utilized for models RBF-M, EXP-M, RBF-A and EXP-A. For models IID-M and IID-A, efficient log-likelihood evaluation is trivially obtained due to the diagonal structure of the covariance matrix.

\begin{table}[htb!]
\centering
\caption{Overview of models used in the case with real-world measurements using multiple sensors. See \autoref{tab:covariance_functions} for the meaning of the abbreviations and the details of the correlation function.}
\label{tab:summary_models_case_2}
\begin{tabular}{@{}llllll@{}}
\toprule
Model    & Shorthand & \begin{tabular}[c]{@{}l@{}}Temporal \\ correlation \end{tabular} & \begin{tabular}[c]{@{}l@{}}Spatial \\ correlation \end{tabular}        & \begin{tabular}[c]{@{}l@{}}Dataset \\ size\tablefootnote{The factor $12$ in the dataset size is included to indicate that each of the six sensors yields two influence lines, one for each controlled loading test as discussed in Section \ref{section:fugro_measurements}.} \end{tabular}             & $\bm{\theta}_\mathrm{c}$         \\ \midrule
$\mathcal{M}_1$    &    IID-M & Independent & Independent        & $193 \times 12$                    & $C_{\mathrm{v}}$, $\sigma_{\mathrm{meas}}$               \\
$\mathcal{M}_2$     &    RBF-M & Radial Basis & Exponential       & $193 \times 12$                  & $C_{\mathrm{v}}$, $\sigma_{\mathrm{meas}}$, $l_{\mathrm{corr,t}}$, $l_{\mathrm{corr,x}}$        \\
$\mathcal{M}_3$     &    EXP-M & Exponential & Exponential           & $193 \times 12$                & $C_{\mathrm{v}}$, $\sigma_{\mathrm{meas}}$, $l_{\mathrm{corr,t}}$, $l_{\mathrm{corr,x}}$          \\
$\mathcal{M}_4$     &    REF-M & Independent & Independent        & $4 \times 12$                    & $C_{\mathrm{v}}$, $\sigma_{\mathrm{meas}}$           \\
$\mathcal{M}_5$    &    IID-A & Independent & Independent        & $193 \times 12$                   & $\sigma_{\mathrm{model}}$               \\
$\mathcal{M}_6$     &    RBF-A & Radial Basis & Exponential       & $193 \times 12$                   & $\sigma_{\mathrm{model}}$, $\sigma_{\mathrm{meas}}$, $l_{\mathrm{corr,t}}$, $l_{\mathrm{corr,x}}$        \\
$\mathcal{M}_7$     &    EXP-A & Exponential & Exponential           & $193 \times 12$                & $\sigma_{\mathrm{model}}$, $\sigma_{\mathrm{meas}}$, $l_{\mathrm{corr,t}}$, $l_{\mathrm{corr,x}}$          \\
$\mathcal{M}_8$     &    REF-A & Independent & Independent        & $4 \times 12$                     & $\sigma_{\mathrm{model}}$               \\ 
\bottomrule
\end{tabular}
\end{table}

\begin{figure}[htb!]
    \centering
    \FIG{\includegraphics[width=0.8\textwidth]{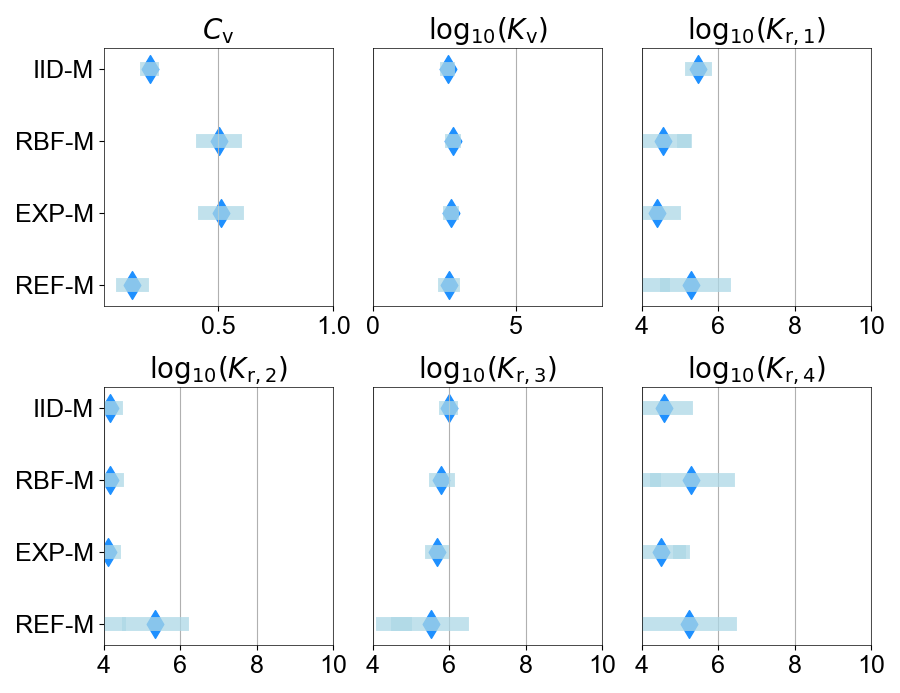}}
    \caption{Comparison of posterior mean and $90\%$ highest density credible intervals for models with multiplicative uncertainty structure.}
    \label{fig:posteriors_case_2b_multiplicative}
\end{figure}

\begin{figure}[htb!]
    \centering
    \FIG{\includegraphics[width=0.8\textwidth]{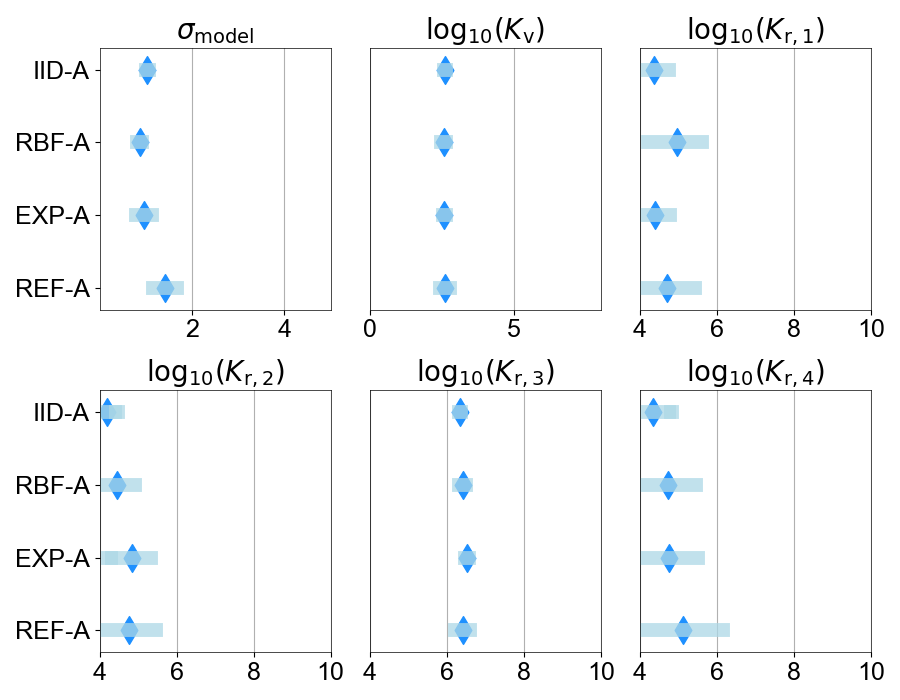}}
    \caption{Comparison of posterior mean and $90\%$ highest density credible intervals for models with additive uncertainty structure.}
    \label{fig:posteriors_case_2b_additive}
\end{figure}

The posterior distribution credible intervals (CI's) for the multiplicative and additive uncertainty models are shown in Figures \ref{fig:posteriors_case_2b_multiplicative} and \ref{fig:posteriors_case_2b_additive} respectively. For the models with additive model prediction uncertainty, it is observed that the reference case REF-A generally results in wider CI's for the inferred parameters, which is expected given the smaller dataset used in this case. Conversely, IID-A typically yields narrower CI's for the physical model parameters. Under a multiplicative model prediction uncertainty, the reference model REF-M yields the lowest point estimate for the COV parameter $C_\mathrm{v}$. Given that only a few hand-selected peaks are considered for the reference models, it is likely that the model prediction uncertainty is underestimated for the REF-M model due to the omission of measurements across the entire length of the bridge, and particularly at locations near the supports where the predicted stress is close to zero. This conclusion is supported by observing that IID-M, while having the same probabilistic model as REF-M and a larger dataset, yields a higher point estimate for $C_{\mathrm{v}}$. Despite the lower $C_\mathrm{v}$, the REF-M model results in wider CI's for the physical model parameters.

The estimated log-evidence for each coupled probabilistic-physical model is provided in \autoref{tab:autogenerated_model_selection_case_2b}. The highest log-evidences are obtained for the models with correlation and additive model prediction uncertainty, RBF-A and EXP-A, while models with multiplicative model prediction uncertainty result in lower evidence, indicating that the mechanisms that contribute to the error between measurements and physical model predictions are closer to being additive in nature for this specific case study. It can also be seen that models where correlation is not considered result in significantly lower evidence. The impact of the probabilistic model and number of parameters on the number of likelihood function evaluations required for convergence is also more pronounced in this case. As shown in the second column of \autoref{tab:autogenerated_model_selection_case_2b}, the EXP and RBF models require approximately two to four times more function evaluations as the corresponding IID models. This highlights the necessity of utilizing an efficient log-likelihood evaluation method in order to reduce the computational cost to a level that makes this approach feasible in practice. It is also emphasized that the applicability of the proposed approach is limited to cases where a computationally cheap physical model or an efficient surrogate model is available.


\begin{table}[htb!] 
        \centering 
        \caption{NFE required for convergence rounded to the nearest thousandth, log-evidence, posterior probability and Bayes factors per model. All models included in the model selection are assumed a priori equally likely.} 
        \label{tab:autogenerated_model_selection_case_2b} 
        \begin{tabular}{llllll}
        \hline
         Model & Shorthand  &    NFE ($\times 1000)$ &   log($\mathcal{Z}$) &  p($\mathcal{M}$) &   Evidence against   \\
        \hline
        $\mathcal{M}_{\mathrm{1}}$ & IID-M &  78 &             -2312.29 &               0.00 & Decisive         \\
        $\mathcal{M}_{\mathrm{2}}$ & RBF-M & 193 &               -94.12 &               0.00 & Decisive         \\
        $\mathcal{M}_{\mathrm{3}}$ & EXP-M & 140 &              -440.28 &               0.00 & Decisive         \\
        $\mathcal{M}_{\mathrm{4}}$ & REF-M & 64 &                -      &               -     &   -         \\
        $\mathcal{M}_{\mathrm{5}}$ & IID-A &  48 &             -3400.96 &               0.00 & Decisive         \\
        $\mathcal{M}_{\mathrm{6}}$ & RBF-A & 204 &              1693.24 &             1.00    & Barely worth mentioning         \\
        $\mathcal{M}_{\mathrm{7}}$ & EXP-A & 196 &              1058.73 &             0.00    & Decisive         \\
        $\mathcal{M}_{\mathrm{8}}$ & REF-A & 30 &                -      &               -     &   -         \\
        \hline
        \end{tabular}
\end{table}

\begin{figure}[htb!]
    \centering
    \FIG{\includegraphics[width=0.9\textwidth]{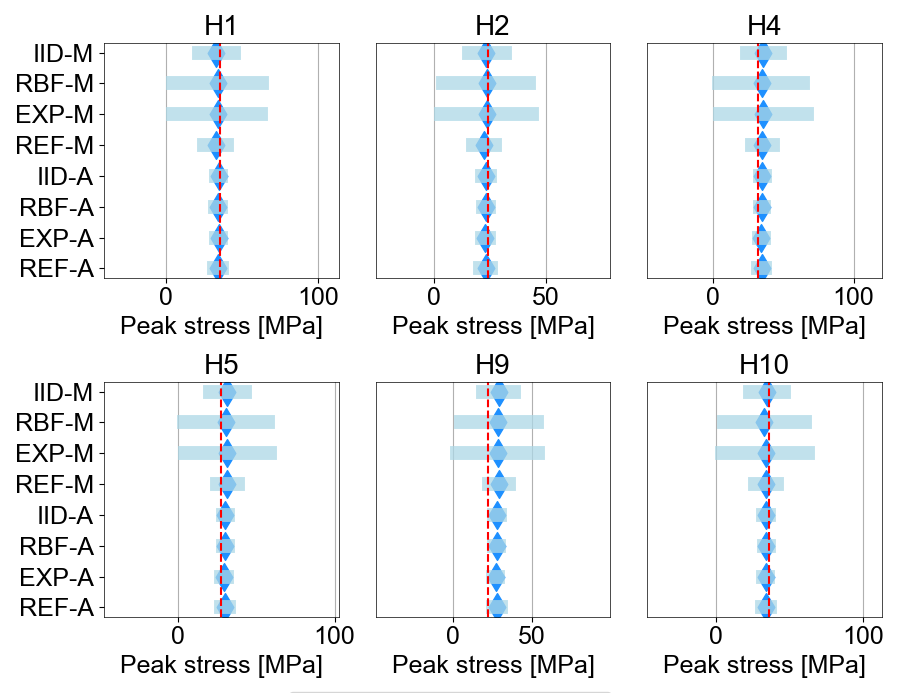}}
    \caption{Comparison of median and $90\%$ credible intervals of the posterior predictive stress distribution per model and sensor at the location of peak stress for a truck on the right lane. The red dashed lines denote the measurements.}
    \label{fig:post_pred_case_2b}
\end{figure}

The degree to which the size of the dataset and the assumptions on the probabilistic model affect the quality of the prediction are reflected in the posterior predictive distribution. For each model, $2000$ samples are drawn from the posterior predictive distribution of the stress influence line. The median and $90\%$ CI's at the location of peak stress for each influence line are plotted in \autoref{fig:post_pred_case_2b}, considering a truck load on the right lane. Significant variations in the widths of the posterior predictive CI's are observed, with the multiplicative models yielding wider CI's compared to the additive models. This may be attributed to the sensitivity of the posterior predictive distributions of the multiplicative models to the uncertainty of the COV parameter $C_{\mathrm{v}}$. Small uncertainties on the posterior of $C_{\mathrm{v}}$ can lead to wide predictive credible intervals at locations where the model response has a large magnitude, such as the peak stress locations shown in \autoref{fig:post_pred_case_2b}. Additionally, it is likely that the multiplicative model may not be an appropriate description of the measured data for the considered case, leading to wide posterior distributions of $C_{\mathrm{v}}$, and consequently wide posterior predictive distributions. This provides additional motivation for performing the model selection procedure described in this study. Furthermore, considering the full dataset and complete independence is shown to result in lower uncertainty in the posterior predictive compared to models where dependence is taken into account for multiplicative models. This is not the case for additive models where no significant differences in the posterior predictive CI's are observed. These results, combined with the calculated evidence and posterior probabilities per model shown in \autoref{tab:autogenerated_model_selection_case_2b} indicate that the assumption of complete independence in the model prediction error can result in overconfident posterior and posterior predictive distributions for certain models.

\section{Conclusions}
\label{section:conclusions}

An approach is proposed to carry out Bayesian system identification of structures when spatial and temporal correlations are present in the model prediction error, and large datasets (in the order of $10^2$ to $10^4$ measurements) are available. To address the issue of the computationally expensive likelihood function, which becomes a bottleneck for large datasets, an approach based on the properties of the exponential kernel function is proposed. It is applicable to both additive and multiplicative model prediction error. Moreover, nested sampling is utilized to compute the evidence under each model and apply Bayesian model selection. The nested sampling method has been shown to be effective for high-dimensional and multi-modal posteriors, indicating that the approach presented in this work is applicable to problems with large numbers of uncertain parameters. Investigating the impact of the physical model parameters on Bayesian inference and model selection could be an interesting avenue for future research.


We conducted a case study using both synthetic and real-world measurements to evaluate the feasibility of performing Bayesian system identification for structures with large datasets. Our proposed approach considers spatial and temporal correlation in the model prediction error. The synthetic example aims to investigate the impact of the dataset size and the assumed structure of the correlation on the inference of the parameters of the probabilistic model, as well as the feasibility of inferring the true probabilistic model from a pool of candidate models, particularly when the number of measurements available is limited. In this example, Bayesian inference and model selection are performed for a set of coupled probabilistic-physical models while refining the spatial and temporal resolution of a dataset. The size of the dataset ranges from $25$ to $2500$ measurements. It is demonstrated that the most likely probabilistic model (which in this case is the a priori known model used to generate the data) can be identified from a pool of candidate models using datasets with as few as $25$ measurements. The size of the dataset is found to have an impact on both the identification of the correct probabilistic model and the accuracy of the point estimates obtained for the uncertain probabilistic model parameters, with larger datasets leading to higher accuracy in both tasks. The structure (e.g. multiplicative or additive) and kernel function considered in the probabilistic model is also found to have a significant influence on the accuracy of the obtained point estimates, particularly for the correlation length parameters. Whereas a relative error below $10\%$ is achieved in the MAP estimates of the probabilistic model parameters of additive models for grids with as few as $10 \times 10$ measurements, grid sizes of $50 \times 50$ or larger are required to obtain similar accuracy with multiplicative models.

Through the IJsselbridge case study, it is demonstrated how Bayesian system identification can be feasibly performed when spatial and temporal dependence might be present. Datasets with up to approximately $2300$ measurements are used to infer the uncertain parameters of the coupled physical-probabilistic models, and Bayesian model selection is applied to determine the most probable model within a pool of candidate models. In addition to updating the physical model and quantifying the uncertainty in the physical model parameters, this approach makes it possible to infer the correlation structure and obtain an improved description of the measurement and model prediction errors. Through the study of the posterior predictive credible intervals and the corresponding model posterior probabilities, both the single and multiple sensor use cases highlight the importance of considering dependencies in the probabilistic model formulation, and demonstrate the benefits of performing Bayesian model selection to determine the posterior probability of different probabilistic models. It is found that the use of a few selected peaks under the i.i.d. assumption for the model prediction error in Bayesian inference can lead to insufficient data and inability to infer all of the parameters of interest, therefore limiting the number and type of parameters that can be identified. Using the full dataset under the assumption of dependence allows for additional parameters to be inferred. In both real-world examples the estimated Bayes factors and posterior probabilities overwhelmingly favour additive error models where correlation is considered, with the EXP-A and RBF-A models obtaining practically $100\%$ posterior probability for the single and multiple sensor cases respectively. 


The proposed approach is based on a multiplicative or additive physical model error. These errors are assumed to follow a Gaussian distribution which determines the likelihood of our model. However, this assumption may not always be realistic in real-world applications. We are currently working on making our model more flexible to better accommodate real-world scenarios.

\begin{Backmatter}

\paragraph{Acknowledgments}
We are grateful to the Dutch Ministry of Public Works and Transport (Rijkswaterstaat) for providing the measurement data used in this study, as well as technical information regarding the IJssel bridge.

\paragraph{Funding Statement}
This publication is part of the project LiveQuay: Live Insights for Bridges and Quay walls (with project number NWA.1431.20.002) of the research programme NWA UrbiQuay which is (partly) financed by the Dutch Research Council (NWO). Part of this research was conducted during the TNO ERP–SI/DT: TNO Early Research Program — Structural Integrity/Digital Twin, use case Steel Bridge. AC gratefully acknowledges the financial support provided by the Alexander von Humboldt Foundation Research Fellowship for Experienced Researchers supporting part of this research.


\paragraph{Competing Interests}
None

\paragraph{Data Availability Statement}
The code used for the synthetic use case is available at \url{https://github.com/JanKoune/bayesian_si_with_correlation}. Restrictions apply to the availability of the real-world measurements, which were used with permission from Rijkswaterstaat.

\paragraph{Ethical Standards}
The research meets all ethical guidelines, including adherence to the legal requirements of the study country.

\paragraph{Author Contributions}
Conceptualization: I.K; A.R; A.S; A.C. Methodology: I.K; A.R; A.S; A.C. Data curation: I.K. Formal Analysis: I.K. Writing original draft: I.K. Supervision: A.R; A.S; A.C. Writing review and editing: A.R; A.S; A.C. All authors approved the final submitted draft.

\begin{appendix}
\section{Sensitivity analysis}
\label{appendix:sensitivity}

For $\mathrm{log}_{10}(K_{\mathrm{r}})$ we assume that any stiffness value from zero (practically pinned support) to infinity (fixed support) is equally likely. In practice the stiffness must be finite and therefore a high value is specified as an upper limit instead. This value is determined by calculating the peak stress predicted at each sensor location as a function of $\mathrm{log}_{10}(K_{\mathrm{r}})$. The upper bound of the support for the prior distribution is chosen as the point where any further increase has a negligible effect on the calculated influence line. The prior of $\mathrm{log}_{10}(K_{\mathrm{v}})$ is determined in a similar manner. All values between zero (practically uncoupled main girders) and infinity (fully coupled main girders) are considered equally likely, and the difference in peak stress for a truck load applied to the left and right lane is calculated. The analyses described previously are performed for the sensors H1, H5 and H10, which are the closest to each of the first three midspans (see \autoref{fig:fugro_sensor_locations}). The results for $\mathrm{log}_{10}(K_{\mathrm{r,4}})$ and $\mathrm{log}_{10}(K_{\mathrm{v}})$ are shown in \autoref{fig:sensitivity_analysis}. 


\begin{figure}[htb!]
	\begin{center}

    \begin{subfigure}[t]{.48\textwidth}
        \begin{center}
            \FIG{\includegraphics[width =\linewidth]{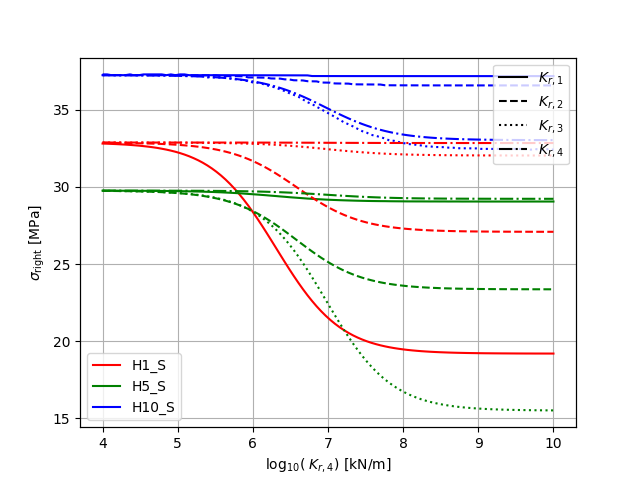}}
            \end{center}
		\end{subfigure}
		\begin{subfigure}[t]{.48\textwidth}
		    \begin{center}
    			\FIG{\includegraphics[width=\linewidth]{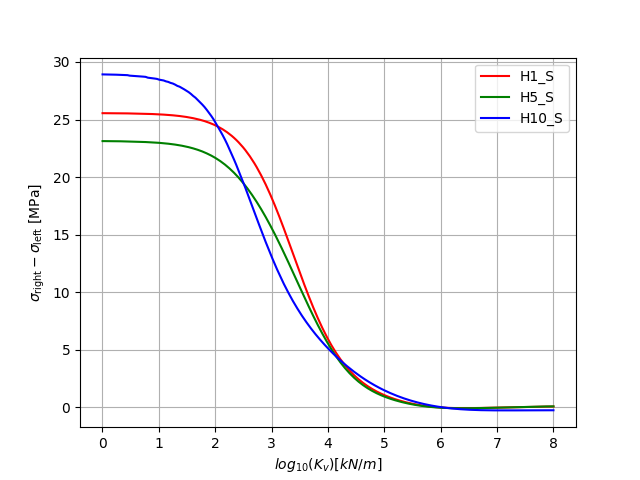}}
  			\end{center}
		\end{subfigure}
		\caption{Peak stress response at selected sensors as a function of $\mathrm{log}_{10}(K_{\mathrm{r}})$ (left) and  $\mathrm{log}_{10}(K_{\mathrm{v}})$ (right).}
  \label{fig:sensitivity_analysis}
	\end{center}
\end{figure}
\section{Measurements}
\label{appendix:measurements}

Measurements are obtained by 34 sensors connected by a total of 9 fiber optic lines to an interrogator sampling at a frequency of $50.0$ Hz. A total of six tests were performed with trucks driving over the left or right lane at a constant speed, with the truck transverse position roughly corresponding to that of the right or left girder depending on the test. Both the transverse position and speed were manually controlled. Different load tests with various truck speeds were performed. A summary of these tests is provided in \autoref{tab:test_times}. In this paper, we only consider the low speed tests (i.e. $20$ km/h) to measure the structural response under approximately static loading conditions. The axle distance and load per axle of the truck used to perform the controlled loading tests are provided in \autoref{tab:truck_fugro}.

\begin{table}[htb!]
\centering
\caption{Controlled loading test parameters.}
\label{tab:test_times}
\begin{tabular}{@{}ccc@{}}
\toprule
Time start {[}CET{]} & Lane  & Speed {[}km/h{]} \\ \midrule
21:56:55             & Right & 20               \\
22:05:55             & Left  & 20               \\
22:21:30             & Left  & 80               \\
22:29:12             & Left  & 80               \\
22:41:25             & Right & 80               \\
22:49:15             & Right & 80               \\ \bottomrule
\end{tabular}
\end{table}

\begin{table}[htb!]
\centering
\caption{Properties of truck used in controlled load tests.}
\label{tab:truck_fugro}
\begin{tabular}{@{}lll@{}}
\toprule
Axle no. & Axle distance {[}m{]} & Load per axle {[}kN{]} \\ \midrule
1        & 2.06                  & 59.35         \\
2        & 1.83                  & 108.82        \\
3        & 1.82                  & 108.82        \\
4        & 1.82                  & 108.82        \\
5        & -                     & 108.82        \\ \bottomrule
\end{tabular}
\end{table}

The truck center of mass is calculated by assuming that the front axle take 12\% of the total load, with the remaining axles taking 22\% of the load. The center of mass is calculated as:

\begin{equation}
    x_{\mathrm{CM}} = \frac{\sum{w_i \cdot x_i}}{\sum{w_i}}
\end{equation}

During processing of the measurement data, it was found that the truck speed deviated from the assumed $20$ km/h and this deviation should be accounted for in the processing. To implement the correction it was assumed that the influence line peak for each sensor occurs when the truck center of mass coincides with the sensor longitudinal position. The time difference $\Delta t$ between the peaks of sensors H1 and H10 was measured. The distance $\Delta x$ between the two sensor positions was then divided by $\Delta t$ to obtain the truck velocity for the left and right lanes equal to $v_l = 21.18$ km/h and $v_r = 21.66$ km/h respectively. The influence lines are obtained by applying a time window to the strain time series. The window start and end times correspond to the first track axle entering the bridge and the last truck axle leaving the bridge respectively, as shown in \autoref{fig:meas_start_end_pos}. The time corresponding to the start and end position can be determined using the known distances $\Delta x_1$ and $\Delta x_2$ and the truck speed calculated previously. A $-0.1$ s shift was applied to the right lane measurements to minimize the discrepancies between the measured and predicted stress influence lines.

\begin{figure}[htb!]
    \centering
    \includegraphics[width=1.0\textwidth]{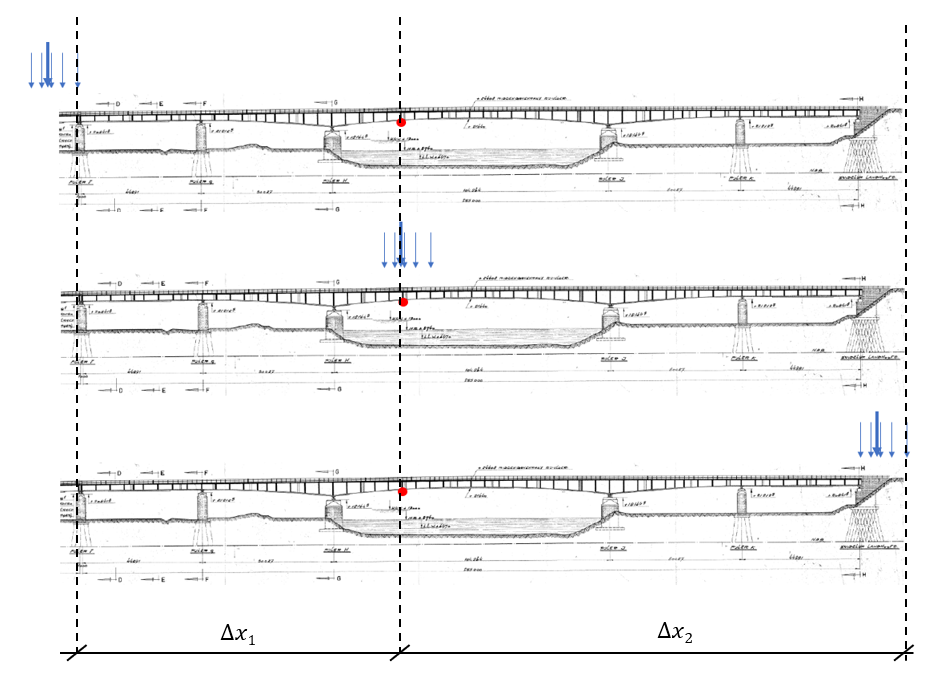}
    \caption{Load position at influence line start (top), peak (middle) and end (bottom).}
    \label{fig:meas_start_end_pos}
\end{figure}
\end{appendix}

\bibliographystyle{apalike}
\bibliography{thesis_references}

\end{Backmatter}

\end{document}